\documentclass{aastex631}

\usepackage[T1]{fontenc}

\usepackage{multirow}
\usepackage{amsmath}
\usepackage{hhline}
\usepackage{array, makecell}
\usepackage{booktabs, tabularx}
\usepackage{comment}

\newcommand{\gray}{$\gamma$-ray}

\newcommand{\soprano}{\texttt{SOPRANO}}

\usepackage{multirow}

\usepackage{xcolor}
\newcounter{tr}
\ifnum \value{tr}>5

\fi

\begin{document}

\title{Modeling blazar broadband emission with convolutional neural networks  - III. proton synchrotron and hybrid models}

\shorttitle{Convolutional neural network for hadronic model}

\shortauthors{Sahakyan et al.}

\author[0000-0003-2011-2731]{N. Sahakyan}
\affiliation{ICRANet-Armenia, Marshall Baghramian Avenue 24a, Yerevan 0019, Armenia}

\author[0000-0003-4477-1846]{D. B\'egu\'e}
\affiliation{Bar Ilan University, Ramat Gan, Israel}

\author[0009-0007-4522-5501]{ A. Casotto}
\affiliation{Chief Scientist, Altair, 640 West California Avenue, Suite 220 Sunnyvale, CA 94086}

\author[0000-0002-8852-7530]{H. Dereli-B\'egu\'e}
\affiliation{Bar Ilan University, Ramat Gan, Israel}

\author[0000-0002-3777-7580]{V. Vardanyan}
\affiliation{ICRANet-Armenia, Marshall Baghramian Avenue 24a, Yerevan 0019, Armenia}

\author[0009-0007-7798-2072]{M. Khachatryan}
\affiliation{ICRANet-Armenia, Marshall Baghramian Avenue 24a, Yerevan 0019, Armenia}

\author[0000-0002-2265-5003]{P. Giommi}
\affiliation{Associated to INAF, Osservatorio Astronomico di Brera, via Brera, 28, I-20121 Milano, Italy}
\affiliation{Center for Astrophysics and Space Science (CASS), New York University Abu Dhabi, PO Box 129188 Abu Dhabi, United Arab Emirates}
\affiliation{Institute for Advanced Study, Technische Universit{\"a}t M{\"u}nchen, Lichtenbergstrasse 2a, D-85748 Garching bei M\"unchen, Germany}

\author[0000-0001-8667-0889]{A. Pe{'}er}
\affiliation{Bar Ilan University, Ramat Gan, Israel}

\begin{abstract}
Modeling the broadband emission of blazars has become increasingly challenging with the advent of multimessenger observations. Building upon previous successes in applying convolutional neural networks (CNNs) to leptonic emission scenarios, we present an efficient CNN-based approach for modeling blazar emission under proton synchrotron and hybrid lepto-hadronic frameworks. Our CNN is trained on extensive numerical simulations generated by \soprano{}, which span a comprehensive parameter space accounting for the injection and all significant cooling processes of electrons and protons. The trained CNN captures complex interactions involving both primary and secondary particles, effectively reproducing electromagnetic and neutrino emissions. This allows for rapid and thorough exploration of the parameter space characteristic of hadronic and hybrid emission scenarios. The effectiveness of the trained CNN is demonstrated through fitting the spectral energy distributions of two prominent blazars, TXS 0506+059 and PKS 0735+178, both associated with IceCube neutrino detections. The modeling is conducted under assumptions of constant neutrino flux across distinct energy ranges, as well as
by adopting a fitting that incorporates the expected neutrino event count through a Poisson likelihood method. The trained CNN is integrated into the Markarian Multiwavelength Data Center (\texttt{MMDC}; \url{www.mmdc.am}), offering a robust tool for the astrophysical community to explore blazar jet physics within a hadronic framework.

\end{abstract}


\section{Introduction}

The detection of the neutrino event IceCube-170922A from the direction of the flaring blazar TXS 0506+056
\citep{IceCube18,2018MNRAS.480..192P}, along with the identification of neutrinos from the same region in the
archival IceCube data \citep{IceCube18b}, marked the beginning of multi-messenger observations of blazars.
Although blazars were historically proposed as potential neutrino-emitting candidates \citep[e.g.,][]{2008PhR...458..173B},
these observations enabled to constrain for the first time the processes
in blazar jets by relying on both electromagnetic and neutrino emissions. Following these observations, several other neutrino events were tentatively associated with blazars owning to their spatial coincidence \citep[e.g.,][]{2020ApJ...899..113P,2020MNRAS.497.2553K,2020ApJ...902...29P, 2020A&A...640L...4G, 2021JCAP...10..082O,2022MNRAS.511.4697P,2022ApJ...932L..25L}, further strengthening the hypothesis that blazars are potential neutrino sources. Among the associated sources, a notable example is PKS 0735+178 \citep[][]{2023MNRAS.519.1396S,2023ApJ...954...70A, 2025A&A...695A.266O, 2024MNRAS.527.8746P}, from which neutrino events were detected by several instruments: IceCube \citep{2021GCN.31191....1I}, Baikal-GVD \citep{Baikal}, the Baksan Underground Scintillation Telescope
\citep{2021ATel15143....1P} and the KM3NeT undersea neutrino detector \citep{2022ATel15290....1F}. Simultaneously to these detections, the source exhibited a high flaring state across all electromagnetic bands \citep[][]{2023MNRAS.519.1396S}, strengthening the connection between neutrino emission and multiwavelength flare of this source.

Blazars are a special class of radio-loud active galactic nuclei characterized by a relativistic jet that is closely aligned with the line of sight of the observer \citep{1995PASP..107..803U}. Due to this small viewing angle, the emission from the blazar jet is strongly Doppler-amplified, making blazars among the most powerful and consistently emitting sources in the extragalactic sky. Blazars are traditionally divided into two subclasses—BL Lacertae objects (BL Lacs) and Flat Spectrum Radio Quasars (FSRQs)—based on the properties of their optical emission lines \citep{1995PASP..107..803U}. FSRQs exhibit strong optical emission lines, whereas BL Lacs are characterized by weak or absent emission lines.

The broadband emission of blazars, extending from radio to the high-energy (HE; $>100$ MeV) and very-high-energy
(VHE; $>100$ GeV) \gray\ bands, exhibits a characteristic double-peaked structure. The first peak, typically located
in the infrared (IR)/optical to X-ray band, is modeled as synchrotron radiation produced by electrons within
the relativistic jet. However, the origin of the second peak remains under debate. It can arise from processes
involving either electrons, within the so-called leptonic scenarios, or protons, within hadronic scenarios.
In leptonic models, the HE/VHE component is attributed to inverse Compton scattering of photons, which can be either
internal (as in the Synchrotron Self-Compton model, SSC) \citep[see, e.g.,][]{1985A&A...146..204G, 1992ApJ...397L...5M, 1996ApJ...461..657B},
or external \citep[External Inverse Compton, EIC, see, e.g.,][]{1994ApJ...421..153S, 1992A&A...256L..27D, 1994ApJS...90..945D, 2000ApJ...545..107B}.

In alternative hadronic models, for which protons are accelerated to high or even ultra-high energies
$ 10^{14} {\rm eV} \lesssim E_p \lesssim 10^{18}$eV,
the HE/VHE component can result from proton synchrotron emission \citep{2001APh....15..121M} or from the emission
of secondaries produced in photo-pion and photo-pair interactions \citep{1993A&A...269...67M, 1989A&A...221..211M, 2001APh....15..121M,
mucke2, 2013ApJ...768...54B, 2015MNRAS.447...36P}. Following the detection of VHE neutrinos from the direction
of several blazars, hadronic models—and their hybrid subtypes (e.g., lepto-hadronic models) are increasingly applied
to model the spectral energy distributions (SEDs) of blazars \citep[see, e.g.,][]{2018ApJ...863L..10A, 2018ApJ...864...84K,
2018ApJ...865..124M, 2018MNRAS.480..192P, 2018ApJ...866..109S, 2019MNRAS.484.2067R, 2019MNRAS.483L..12C, 2019A&A...622A.144S,
2019NatAs...3...88G, GBS22, 2023MNRAS.519.1396S, RPG24,RKP24}.

In \citet{BSD23} and \citet{SBC24}, we highlighted the challenges of modeling the SEDs of blazars
and interpreting their multi-wavelength spectra using leptonic scenarios when time-dependent codes are
coupled with Bayesian fitting tools. In this context, hadronic models present even greater challenges due to (i)
the complexity and variety of particle interactions involved in the computations and (ii) large variations in the
calculated SEDs when the model transitions between different dominant emission processes. This complexity makes it
practically impossible to make a fit with hadronic models unless significant simplifying assumptions are made
about various parameters or the model setups.
However, even under such assumptions, considerable computational resources are required. In addition, the
computational results are hardly reusable; for each new dataset, the computations must be repeated.
Given the growing number of blazars associated with VHE neutrinos, performing comprehensive modeling of the data and
fully leveraging the potential of multimessenger data is becoming increasingly important. Consequently, the ability
to fit hadronic models by taking into account both constraints from electromagnetic and neutrino
observations is becoming essential.

This study is the third contribution in the project called "Modeling Blazar Broadband Emission with Convolutional Neural Networks."
The primary goal of this project is to transform the complex and time-consuming process of calculating the numerical models
into an efficient and fast task that requires minimal computational resources, leveraging the advantages provided by neural
networks, in order to enable computational resources to be diverted to the fitting process itself. In the first two
papers of the series, \citet{BSD23} and \citet{SBC24}, the neural network approach successfully
transformed the numerically expensive self-consistent SSC and EIC models calculation into rapid models. All the models calculated for the project "Modeling Blazar Broadband Emission with Convolutional Neural Networks." are made accessible online via the Markarian Multiwavelength Data Center \citep[\texttt{MMDC};
\url{www.mmdc.am}][]{2024AJ....168..289S}. In this paper, we extend the methodology to a
multi-messenger model, namely the so-called hadronic model. The neural network was trained on the
numerical output obtained by \soprano{} \citep{GBS22}. We demonstrate here that the resulting neural
network accurately reproduces both electromagnetic and neutrino emissions while accounting for the relevant cooling processes
for primary and secondary particles. We further demonstrate that the network, when coupled with a Bayesian fitting engine, enables a comprehensive
exploration of the parameters in hadronic models, representing a significant step forward in modeling multimessenger data
from blazar observations.

The creation of surrogate models is becoming
increasingly common and popular in the field of high-energy astrophysics. For blazar SED modeling, in addition to our previous works \citep{BSD23, SBC24}, \citet{TVP24} trained a neural
network to create a surrogate model of an SSC model. In the context of gamma-ray bursts, \citet{BvL23} emulated a multiwavelength afterglow
and \citet{WS25} created a neural network to approximate the joint emission of an afterglow and a kilonova emission.
In all cases, the resulting models are sufficiently fast and accurate to enable detailed parameter exploration and fits.

The paper is structured as follows: the model is described in Section \ref{sec:model}, the computation of synthetic data,
training of the convolutional neural network (CNN), and its validation are presented in Section \ref{sec:cnn}. In Section
\ref{sec:appl}, the trained CNN is applied to model the multimessenger SEDs of TXS 0506+059 and PKS 0735+178, two
sources which have been associated to neutrino emission. The availability of the trained model on the \texttt{MMDC} platform
is discussed in Section \ref{sec:aval}, and the conclusion is presented in Section \ref{sec:conc}.

\section{A hadronic model including both proton synchrotron and hybrid models}\label{sec:model}

In this study, we further develop the model applied in \citet{BSD23} and \citet{SBC24} to analyze the
multimessenger data observed from blazars. Previous models explaining the broadband emission from BL Lacs and FSRQs considered the acceleration and injection of electrons only, discarding the possible
acceleration of protons, if present in the jet. If protons can be accelerated
alongside electrons, their emission could
contribute to the observed HE and VHE \gray\ data. We note that the efficient acceleration of protons to highly relativistic energies $E_p \gtrsim 10^{16}$eV within the jet is a requirement for the production of VHE neutrinos within the jet. The advantage of modeling multimessenger data lies in its ability to estimate the content of particles (electrons and protons) and magnetic fields, which is crucial for understanding and constraining these systems.

In the model under consideration, we assume that both electrons and protons are accelerated
within the jet and injected into an emitting region, which is assumed to be a sphere of radius $R$.
The emission region is filled with a tangled and uniform magnetic field of strength $B$ and moves along the jet at
relativistic speed toward the observer with a bulk Lorentz factor $\Gamma$, assumed to be equal to the Doppler factor
$\delta$ under the condition of a small jet inclination angle. The particles (electrons and protons) are injected into
the emitting region following a power-law with an exponential cut-off distribution given by:
\begin{equation}
Q^{\prime}_{\rm i}(\gamma_{\rm i})= \left \{ \begin{aligned}
& Q^{\prime}_{0,\rm i} \gamma^{-p_{\rm i}}_{\rm i} \exp \left ( - \frac{\gamma_{\rm i}}{\gamma_{\rm i, max}} \right ) & ~~~~ & \gamma_{\rm i, min} \leq \gamma_{\rm i} ,  \label{dist} \\
& 0 & & {\rm otherwise,}
\end{aligned} \right.
\end{equation}
where $i = e, p$ represents electrons and protons, respectively, $p_{\rm e}$ denotes the power-law index of the injected electrons, $\gamma_{\rm e, min}$ is the minimum Lorentz factor of the electrons and $\gamma_{\rm e, max}$ is the maximum Lorentz factor, beyond which acceleration becomes inefficient. Similarly, $p_{\rm p}$, $\gamma_{\rm p, min}$ and $\gamma_{\rm p, max}$ correspond to the power-law index, minimum and maximum Lorentz factor of the injected protons, respectively. The normalization factor for the injection of electrons and protons, \( Q^{\prime}_{0,\rm i} \), is determined from the assumed particle luminosities:
\begin{equation}
L_{\rm i, jet} = \pi R^2 \delta^2 m_{\rm i} c^3 \int_1^{\infty} \gamma_{\rm i} Q_{\rm i}^{'} d\gamma_{\rm i},
\end{equation}
where \( m_{\rm i} \) is the mass of the electrons or protons.

The injected particles interact within the emitting region, producing photons (the electromagnetic component) and neutrinos
(the multi-messenger component). The temporal evolution of the injected particles defines their spectrum
and is used to compute their radiative signature. For completeness, we provide a short description of the kinetic
equations describing the evolution of the system and refer the reader to \citet{GBS22} in which we provide the detailed expressions
for the different interaction and cooling rates, and how they are numerically treated in \soprano{}. The Fokker-Planck diffusion equation is used to determine the temporal
evolution of the electron and proton distributions, accounting for all relevant cooling processes. The photon
evolution is described by an integro-differential equation. The kinetic equations for the electrons and photons are:
\begin{align}
\left \{  \begin{aligned}
    \frac{\partial N_e}{\partial t} (\gamma_{\rm e} ) &= \frac{N_e}{t_{\rm esc}} + \frac{\partial}{\partial \gamma_{\rm e} } \left [  ( C_{\rm IC} + C_{\rm sync}  ) N_e \right ] + Q_{\gamma \gamma \rightarrow e^+e^- } +Q_{p} ,    \\
    \frac{\partial N_\gamma}{\partial t} (x) &= \frac{N_\gamma}{t_{\rm esc}} + Q_{\rm  sync} + R_{\rm IC} N_\gamma - S_{\gamma \gamma \rightarrow e^+e^-}, 
\end{aligned} \right. \label{eq:kinetic_equation}
\end{align}
where $N_\gamma$ and $N_e$ are the distribution functions of photons and electrons, respectively, $x$ is the photon
energy, and the escape time is set to $t_{\rm esc} = t_{\rm dyn} = R/c$. The coefficients $C_{\rm IC}$, which
depends on the photon distribution $N_\gamma$, and $C_{\rm sync}$
represent the cooling due to inverse Compton and synchrotron processes, respectively. In addition, $Q_{\gamma \gamma \rightarrow e^+e^- }$
and $S_{\gamma \gamma \rightarrow e^+e^-}$ are the source and sink terms related to pair production, while $R_{\rm IC}$
is the redistribution kernel for Compton scattering. Note that the sink terms $S$ appear with a minus sign as particles
are removed from the system by this process. Finally, $Q_p$ represents the source of all electrons produced in hadronic channels,
in particular via the photopair (Bethe-Heitler) process and muon decay.

Similarly, the temporal evolution of the proton
distribution function is determined through the equation
\begin{equation}
        \frac{\partial N_p}{\partial t} (\gamma_p) = C_{p\gamma \rightarrow p \pi} + C_{p\gamma \rightarrow e^+ e^-} + \frac{\partial}{\partial \gamma_p} \left [ C_{\rm synch} N_p \right ] - S_{\gamma p \rightarrow n \pi } + Q_{\gamma n \rightarrow p \pi} + \frac{N_p}{t_{\rm esc}}.
    \label{eq:p}
\end{equation}
Here, $C_{p\gamma \rightarrow e^+ e^-}$, $C_{p\gamma \rightarrow p \pi}$, and
$C_{\rm synch}$ represent proton cooling through photo-pair and photo-pion interactions and synchrotron cooling, respectively.
Protons are produced through photohadronic interactions between photons and neutrons at a rate $Q_{\gamma n \rightarrow p \pi}$
and converted to neutrons for a substantial fraction of photopion interactions at a rate $S_{\gamma p \rightarrow n \pi }$.
The temporal evolution of all produced secondaries is similarly described through kinetic equations, including synchrotron cooling
for muons and charged pions. High energy neutrinos are produced in charged pions and muons decays. For the complete list of
kinetic equations and the processes considered, see \citet{GBS22}.

The differential equations presented in Equations \ref{eq:kinetic_equation} and \ref{eq:p} are solved using
\soprano{} \citep{GBS22}. \soprano{} is an implicit kinetic code that follows the time evolution of
electrons/positrons, photons, neutrons, and the secondaries produced in photo-pion and photo-pair interactions,
enabling a time-dependent interpretation of observed data. \soprano{} has been used to study multi-messenger emission
from several well-known blazars (TXS 0506+059 \citep{GBS22}, Mrk 501 \citep{2023ApJS..266...37A}, PKS 0735+178
\citep{2023MNRAS.519.1396S}), as well as to generate the training sample used for the creation of the surrogate models for the SSC and EIC processes \citep{BSD23, SBC24}.

The differential equations are solved under the assumption that the system achieves equilibrium, where its evolution becomes time-independent. By evolving the system until \( t = 4t_{\text{dyn}} \), the variation in the spectrum becomes negligible, indicating that the system has reached equilibrium. In the computations, the blob radius is assumed to remain constant, with no expansion, and adiabatic losses are ignored. The output generated by \soprano{}  represents the multi-messenger signature of particle interactions inside the jet and serves as a model database for the training set.

\section{Numerical Model: Computation, CNN and Validation}

\label{sec:cnn}

In this section, we discuss the parameter range, sampling, computation of the SEDs, training of the CNN, and
validation. Unlike the SSC \citep{BSD23} and EIC \citep{SBC24} models for which respectively $2 \times 10^5$
and $10^6$ spectra were generated for the training set, the hadronic model is more complex, harbors fast changes
with respect for variation of the input parameters, and therefore required a larger number of spectra. In order to balance model accuracy and calculation costs, we performed the computation of $7 \times 10^6$
spectra, specifically distributed in the regions of interest of the parameter space. 

\subsection{Parameter Ranges and Sampling}
\begin{table*}
\centering
\begin{tabular}{lccccl}
    \hline
     Parameter & Units  & Symbol  & Minimum & Maximum & Type of distribution \\ \hhline{|=|=|=|=|=|=|}
     Doppler boost & -  & $\delta$ & 3.5 & 100 & Linear \\ 
     Blob radius        & cm &     R    &  $10^{14.5}$   &  $10^{18}$  &  Logarithmic \\
     Minimum electron injection Lorentz factor & - & $\gamma_{\rm e,min}$ & $10^{1.5}$ &  $10^5$ &  Logarithmic \\
     Maximum electron injection Lorentz factor & - & $\gamma_{\rm e, max}$ & $10^2$ &  $10^8$ & Logarithmic \\
     Maximum proton injection Lorentz factor & - & $\gamma_{\rm p,max}$ & $10^3$ &  $10^{11}$ & Logarithmic \\
     Injection index electrons & - & $p_e$ & $1.7$ & 5 & Linear \\
     Injection index protons & - & $p_p$ & $1.6$ & 3.5 & Linear \\
     Electron luminosity & erg.s$^{-1}$ & $L_e$ & $10^{42}$ & $10^{49}$ &  Logarithmic \\
     Proton luminosity & erg.s$^{-1}$ & $L_p$ & $10^{42}$ & $10^{52}$ &  Logarithmic \\
     Magnetic field  &  G  & $B$  & $10^{-3}$  & $10^{3.5}$ & Logarithmic  \\ \hline
\end{tabular}
    \caption{Description of the dataset used for training the CNN, showing the unit and
    symbol associated with each parameter, as well as their respective ranges and the
    distribution of values across discrete parameters. The large spam of the parameters,
    which might look like unrealistic for blazar physics, allows for a thorough parameter
    exploration and unbiased fits. The dataset comprises $3\times10^6$
    individual spectra uniformly distributed in the full domain, an additional $2\times 10^6$ spectra in the region of the P-sync model
    as well as $2\times 10^6$ in the parameter space of the hybrid model. In total, the model is trained on $7\times 10^6$
    spectra calculated with \soprano{}.}
    \label{tab:table_parameters}
\end{table*}

Our analysis does not specialize to a specific hadronic scenario and considers simultaneously both \textit{(i)} the proton-synchrotron model (hereinafter P-syn) for which the HE emission is primarily produced by proton synchrotron radiation, and \textit{(ii)} in the hybrid model for which the low- and high-energy peaks are explained by leptonic processes, while the proton content in the jet is constrained by X-ray radiation from secondaries produced via Bethe–Heitler and photo-pion processes. We considered a very wide range of parameters, allowing both scenarios to be represented at once. This also allows the fit to decide which scenario best describes the data without having to perform model comparison.

In this hadronic model, there are 10 free parameters: the power-law indices
of the electrons and protons $p_e$ and $p_p$, respectively, the minimum and maximum Lorentz factors of the electrons,
$\gamma_{\rm e,min}$ and $\gamma_{\rm e,max}$, respectively, the maximum Lorentz factor of the protons $\gamma_{\rm p,max}$\footnote{
In the current model, the minimum Lorentz factor of the protons is set to $\gamma_{\rm p,min}=1.2$.}, the injection
luminosities of the electrons ($L_e$) and protons ($L_p$), the Doppler factor $\delta$, the emission
region radius $R$, and the magnetic field B in the emission region. The units and the ranges of these parameters are provided
in Table \ref{tab:table_parameters}.

A wide range of parameters for both electrons and protons is considered.
For example, the power-law index of the electrons is linearly sampled in the range $1.7 \leq p_e \leq 5.0$, while the proton's spectral index
is sampled in the range $1.6 \leq p_p \leq 3.5$. Although acceleration theories disfavor steep indices ($p > 3.5$), we consider
even steeper injection spectra for electrons to comprehensively investigate the parameter space without being constrained by its
boundaries. The characteristic Lorentz factor of the electron and proton distribution functions are linearly sampled in the intervals
$1.5 \leq \log(\gamma_{\rm e,min}) \leq 5$, $2 \leq \log(\gamma_{\rm e,max}) \leq 8$ and $2 \leq \log(\gamma_{\rm p,max}) \leq 11$.
This large proton Lorentz factor enable the consideration of ultra-high-energy protons. The injection luminosities of both electrons and protons are also logarithmically sampled over a very wide range: $42 \leq \log(L_e) \leq 49$ and $42 \leq \log(L_p) \leq 52$, respectively. This allows for a comprehensive exploration of all possibilities to explain the SED, namely either (1) leptonic processes explain the full SED or (2) synchrotron radiation from protons and secondaries from $p\gamma$ interactions explains the HE component, while the LE component is produced by synchrotron radiation from electrons. With this method, the multi-wavelength dataset defines the preferred model, alongside neutrino constraints that can only be produced for a high proton luminosity. Based on experience gained from modeling SSC and EIC scenarios, the Doppler  factor (\(\delta\)) is linearly sampled in the range $3.5 \leq \delta \leq 100$. Although very high Doppler factors are physically unrealistic for blazars,  such values are included to accommodate a fair sampling of the posterior distributions of this parameter. In principle, models with a high neutrino yield require a large compactness, which can be achieved by a small $\delta$. The magnetic field $B$ and the emission region radius $R$ are logarithmically sampled in the ranges $-3 \leq \log(B) \leq 3.5$ and $14.5 \leq \log(R) \leq 18$, respectively. By using these parameter ranges, we generated the spectra for $3 \times 10^6$ parameter sets.

Spectral variations can be large with respect to small changes in the value of some parameters, particularly in regions where the
dominant process changes. To improve the accuracy of the model, we generated
additional spectra focused on the parameter space relevant for P-syn and hybrid models. Specifically, for the hybrid
scenario, we considered magnetic field strength  between $10^{-2.5}$ and $10^{1}$G, and proton Lorentz factor in the range $10^{4}$ to $10^{7}$,
while keeping all other parameters as in Table \ref{tab:table_parameters}. For the P-syn model, we used
$B$ between $10^{-1}$ and $10^{2.5}$G and $\log(\gamma_{\rm p,max})$ between ${8}$ and ${10.5}$.
For each of these two subsets, $2\times 10^6$ spectra were generated,  and were merged with those from the full parameter
space to improve the accuracy of the network in these regions. In total, the database on which the convolutional neural network is trained consists of $7 \times 10^6$ spectra, whose parameters are distributed to optimize the precision of the network
for both P-syn and hybrid models.

The parameters are sampled from the domain with
the ronswanson library \citep{Bur23}. Instead of performing regular sampling of the parameter space, which would require the computation
of a prohibitively large number of SEDs, we employ the Latin hypercube sampling method \citep[see][]{MBC00,Via16}. This approach ensures
a comprehensive, uniform, and representative coverage of the multidimensional parameter space, making it an efficient choice for
handling high-dimensional models.

\subsection{ Computation of the Spectra on the Aznavour Supercomputer}

The total computational workload involves generating $7 \times 10^6$ spectra where each spectrum is computed independently. For the hadronic model discussed in this paper, the computation of a single SED {with \soprano{} requires
8 cores for about 1 minute, making it practically impossible to execute the computation
of the entire dataset on a standard server. For this reason, the “Aznavour” supercomputer was used, which  consists of Bullx BL720 nodes,
each equipped with two E5-2698V4 processors and 64 GB of DDR4 memory per node. Another challenge is managing
$7 \times 10^6$ jobs effectively. To address this, the FNC workload manager\footnote{For more information, see the website of Altair:
\url{https://altair.com/newsroom/articles/phoenix-rising-next-generation-job-scheduling-with-fenice}} developed by Altair Labs was used.
Compared to conventional schedulers, FNC offers several advantages: it can efficiently manage millions of jobs, retains the state of
every job even after completion, includes a "Rapid Scaling" module that dynamically adjusts the workload to match available resources,
and provides a browser-based interface for easy workload monitoring and result inspection. These features make FNC an ideal solution
for large-scale computational management. 

\subsection{Convolutional Neural Network}

To construct our surrogate model, we adopted the same CNN architecture presented in \citet{BSD23}, and
further used in \citet{SBC24}. Given the large spectral variability for the hadronic model
compared to the SSC or EC models, and to ensure that the CNN adequately captures the full range of spectral features,
we divided the input parameter space into four distinct regions:
\begin{itemize}
\item Region 1:  $ 42 < \log_{10}(L_e) < 46.5 $   and $ 42 < \log_{10}(L_p) < 48 $,
\item Region 2:  $ 42 < \log_{10}(L_e) < 46.5 $   and $ 46 < \log_{10}(L_p) < 52 $,
\item Region 3:  $ 44.5 < \log_{10}(L_e) < 49 $   and $ 42 < \log_{10}(L_p) < 48 $,
\item Region 4:  $ 44.5 < \log_{10}(L_e) < 49 $   and $ 46 < \log_{10}(L_p) < 52 $.
\end{itemize}
We train a separate CNN within each of the four parameter space regions. These regions are designed to overlap substantially to avoid boundary effects when applying the surrogate model. Specifically, the lower (upper) boundary of each region is extended upward (downward) by half an order of magnitude. In the overlapping regions, the outputs of the corresponding networks are averaged to produce a smooth prediction. We note that, in the case of neutrino spectra, the variation is less pronounced, and thus there is no need to partition the parameter space; a single network is trained instead.

For each of the four parameter space regions described above, the training sample preparation proceeds as follows.
We first removed spectra for which the numerical calculation failed at any point during the Newton-Raphson iterations
used for the implicit integration of the discretized kinetic equations. For each valid spectrum, data points lying
fifty orders of magnitude below the spectral power peak were replaced with a constant value equal to the peak flux
divided by $10^{50}$. The spectral data were then logarithmically transformed and rescaled to lie within the interval
$[-1, 1]$. To suppress spurious oscillations in the network outputs, we adopted the approach introduced in \citet{BSD23},
training the network not only on the target spectral values but also on three finite-difference approximations of the
spectral derivative. The input parameters were independently rescaled to the interval $[-1, 1]$, without any additional
transformation. The dataset was randomly split into a training set (80\%), a validation set (10\%), and a test set (10\%).
We employed the L1 loss function, corresponding to the mean absolute error, and optimized the network using the Adam optimizer.
The learning rate was scheduled to decay from $10^{-3}$ to $10^{-5}$, passing through $10^{-4}$, with transitions at
epochs 30 and 60 during training. For the neutrino sector, we used a single network for each neutrino species for the full
parameter space. The cleaning, normalization and training procedures are the same as for the photons.

Once trained, the surrogate model, composed of 4 CNN in each photon regions and 2 CNN for the neutrinos,
is sufficiently accurate to be used in parameter explorations and fits of the multi-wavelength and multi-messenger
data. In Figure \ref{fig:CNN_performance}, we show a comparison between CNN results and training and validation
databases. Each line in this figure corresponds to one of the different spectral regions. Clearly, the models
performed as expected. This accuracy is attested by several performance metrics which we only report for the model together with the derivatives. Across the 4 photons models, the average $R^2$ score is 0.66, the mean squared error (MSE) is $2.25 \times 10^{-3}$ and the mean
absolute error (MAE) is $5.05 \times 10^{-3}$. Similar results hold for neutrinos. 

\begin{figure}[h]
\centering
\includegraphics[width=0.8\textwidth]{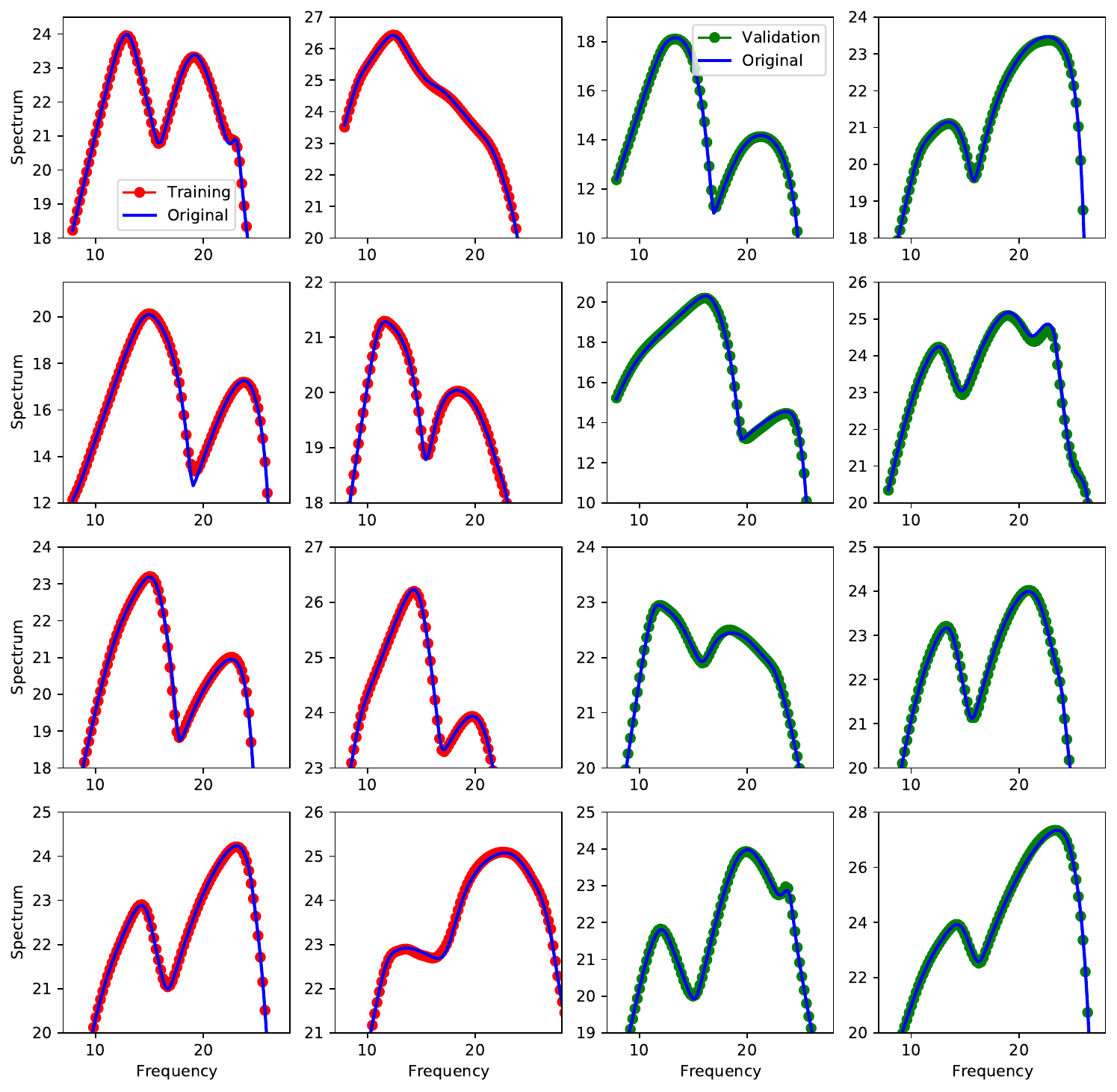}
\label{fig:CNN_performance}
\caption{Comparison between spectra (arbitrary unit) calculated by the CNN and from the training database (two left-most columns) or the validation database (two right-most columns). Each line corresponds to a different parameter space. As can be seen from this figure, the CNN results matches those from the database, demonstrating that this approach produces results with a great accuracy. }
\end{figure}

\section{Applications: Modeling Multimessenger Emission of TXS 0506+059 and PKS 0735+178}\label{sec:appl}

\begin{table}
    \centering
\caption{Model parameters for TXS 0506+059 and PKS 0735+178. The first and second column presents the 
best-fit parameters for TXS 0506+059 with a Gaussian and a Poisson likelihood for the neutrino respectively. The third column lists the best-fit parameters for
PKS 0735+178 assuming a Poisson likelihood for the neutrino. For both fits with a Poisson likelihood, we assume
the detection of a single neutrino detection by IceCube over a one-year long exposure.}
\label{tab:param}
    \begin{tabular}{c||c||c||c}
    \hline
    Parameters  &  \multicolumn{2}{c||}{TXS 0506+059}  & PKS 0735+178 \\\hline 
    Neutrino Likelihood  &  Gaussian  &  Poisson  & Poisson \\\hline
    Preferred model type & Hybrid & P-syn  & P-syn/Hybrid \\\hline 
    \( p_e \) & \( 1.77 \pm 0.05 \) & \( 2.87 \pm 0.39 \) & \( 2.62 \pm 0.45 \) \\
    \( p_p \) & \( 2.03 \pm 0.17 \) & \( 2.00 \pm 0.13 \) & \( 1.87 \pm 0.21 \) \\
    \( \log_{10}(\gamma_{\rm e,max}) \) & \( 4.76 \pm 0.25 \) & \( 3.25 \pm 1.19 \) & \( 3.10 \pm 0.68 \) \\
    \( \log_{10}(\gamma_{\rm e,min}) \) & \( 1.52 \pm 0.12 \) & \( 2.0 \) & \( 2.0 \) \\
    \( \log_{10}(\gamma_{\rm p,max}) \) & \( 6.37 \pm 0.29 \) & \( 8.00 \pm 0.30 \) & \( 7.95 \pm 1.15 \) \\
    \( \delta \) & \( 15.49 \pm 17.43 \) & \( 23.18 \pm 10.64 \) & \( 26.00 \pm 23.13 \) \\
    \( \log_{10}(B/[\rm G]) \) & \( -1.52 \pm 0.29 \) & \( 2.73 \pm 0.62 \) & \( 2.88 \pm 1.59 \) \\
    \( \log_{10}(R/[\rm cm]) \) & \( 17.76 \pm 0.64 \) & \( 14.53 \pm 0.36 \) & \( 14.55 \pm 0.89 \) \\
    ~~~\( \log_{10}(L_{\rm e}/[\rm erg\:s^{-1}]) \)~~~ & ~~~ \( 46.50 \pm 0.20 \) ~~~ & ~~~ \( 43.99 \pm 0.29 \) ~~~ & ~~~ \( 44.61 \pm 0.89 \) ~~~ \\
    ~~~\( \log_{10}(L_{\rm p}/[\rm erg\:s^{-1}]) \)~~~ & ~~~ \( 51.18 \pm 0.38 \) ~~~ & ~~~ \( 46.69 \pm 0.55 \) ~~~ & ~~~ \( 46.91 \pm 1.60 \) ~~~ \\\hline
    \(\log_{10}(L_{\rm B}/[\rm erg\:s^{-1}]) \) & \( 44.43 \) & \( 46.82 \) & \( 47.26 \) \\
    \hline
  \end{tabular}
\end{table}

The trained CNN is now applied to model multimessenger data from blazar observations. The best
candidates for such applications include TXS 0506+059, associated with the IceCube-170922A event, and PKS 0735+178,
located in the sky direction of the IceCube-211208A event. The fit determines the parameters of the emitting electrons — $p_e$,
$L_e$, $\gamma_{\rm e,min}$ and $\gamma_{\rm e,max}$—as well as the parameters of the protons — $p_p$, $L_p$, $\gamma_{\rm p,min}$,
and $\gamma_{\rm p,max}$. The fit also constrains the parameters of the emitting region, including the Doppler factor
$\delta$, magnetic field $B$, and size $R$.

\subsection{Likelihood and sampling}
\label{subsec:likelihood}

The posterior distributions are sampled with
the MultiNest algorithm \citep{FHB09}, with 1000 active points and a tolerance set to 0.5. We use non-informative flat priors.
For the electromagnetic data, we selected a Gaussian likelihood and deferred the treatment of Poisson statistics and instrument response
functions to future works. This task promises to be particularly important at the HEs, from the X-ray band (Swift,
XMM-Newton, etc.) to observations by Fermi LAT and Cherenkov Telescope Array Observatory (CTAO) as it would provide data consistency and information that the Gaussian likelihood cannot capture.
We note that the absorption due to the extragalactic background light (EBL) is accounted for using the model by \citet{2011MNRAS.410.2556D}. The correction is performed by multiplying the CNN results by the absorption.

For the neutrino likelihood, we adopted two approaches to illustrate how neutrino data can help constrain the fit.
First, we considered a Gaussian likelihood similar to that of the electromagnetic sector. This requires the neutrino
flux to be given in at least one energy band. We found that a "single" energy band is not constraining enough and
in the rest of this analysis, we instead used two flux points, see below for details. Alternatively, we implemented a Poisson
likelihood \citep{Cash79}
\begin{align}
    \log_{10} \mathcal{L} = 2 \sum_{i=1}^n (t m_i) - S_i \log_{10} \left ( t m_i \right ) + \log_{10} \left (  S_i ! \right )
\end{align}
where the sum runs over all energy bins $i$, $t$ is the observation time, $m_i$ is the neutrino rate predicted by the model
in energy bin $i$ and $S_i$ is the observed number of counts in this bin. In practice, here, we only consider one energy bin,
which covers the full effective area of IceCube taken from \citet{2019ICRC...36.1021B}. The neutrino flux predicted by the
CNN is multiplied by the differential effective area of IceCube and integrated over energy to calculate $m_i$. This
method quantifies the probability of detecting neutrino events with IceCube while considering the constraints imposed by
the electromagnetic spectrum. In addition, neutrino oscillations is taken into account within the quasi-two neutrino oscillation assumption. The number of muon neutrinos arriving on Earth is given by 
\begin{align}
    N_{\nu_\mu} = 0.575 N_{\nu_\nu}^{s} + 0.425 N_{\nu_e}^s
\end{align}
where the superscript $s$ indicates source quantities, before the propagation from the source to Earth \citep{FGV18}.

\subsection{TXS 0506+059}
TXS 0506+056, a BL Lac-type blazar located at a redshift of $z = 0.336$ \citep{2018ApJ...854L..32P}, is the first blazar identified with neutrinos spatially coincident with its sky direction. On 22 September 2017, the IceCube Neutrino Observatory detected a VHE neutrino event, IceCube-170922A, with an energy of approximately 290 TeV, coincident in direction and time with a \gray\ flare from TXS 0506+056 \citep{IceCube18}. Furthermore, analysis of 9.5 years of IceCube archival data revealed an excess of $13 \pm 5$ neutrino events between September 2014 and March 2015 from the same direction, providing $3.5\sigma$ evidence for prior neutrino emission independent of the 2017 event \citep{IceCube18b}. A detailed analysis of the spatial, temporal, and energy characteristics of the region around IceCube-170922A established TXS 0506+056 as the sole counterpart to these neutrino emissions \citep[e.g.,][]{2018MNRAS.480..192P}. Consequently, the multimessenger emission from TXS 0506+056 has been extensively modeled in various scenarios \citep[see, e.g.,][]{2018ApJ...863L..10A, 2018ApJ...864...84K, 2018ApJ...865..124M, 2018MNRAS.480..192P, 2018ApJ...866..109S, 2019MNRAS.484.2067R, 2019MNRAS.483L..12C, 2019A&A...622A.144S, 2019NatAs...3...88G, GBS22}.

Here, we model the SED of TXS 0506+056 during the period when the
IceCube-170922A event was observed, using data gathered by \citet{IceCube18}. Although
the SED is sufficiently well sampled across the electro-magnetic spectrum to reconstruct both the low- and high-energy emission components, the lack of observations between $10^{12}$ and $10^{14}$ Hz introduces uncertainties in reliably estimating the electron minimum Lorentz factor $\gamma_{\rm e,min}$. 
Allowing $\gamma_{\rm e,min}$ to vary freely across the full range used in model training
(see Table \ref{tab:table_parameters}) results in an excessively steep electron injection
spectrum and a very large $\gamma_{\rm e, min}$. Therefore, $\log_{10}(\gamma_{\rm e,min})$
was limited to vary in the range 1.5–2.0. Following the approach of \citet{BSD23} and
\citet{SBC24}, only data above $10^{12}$ Hz were included in the fit,
as the radio band emission is expected to be self-absorbed and likely originates from a distinct, extended
region of the jet.

To perform the fit with multi-messenger constraints, information on the neutrino spectrum is required.
We treat the neutrino spectrum with both methods detailed in section
\ref{subsec:likelihood}, starting with the Gaussian likelihood.
In this case, we find that only providing a single measurement at the most
likely energy of IceCube-170922A is insufficient,
as many models could reproduce this single data point at that specific energy.
Additionally, the flux normalization of the neutrino event remains uncertain. The IceCube Collaboration reported
upper limits on the neutrino flux corresponding to the detection of one IceCube-170922A–like event over exposure
periods of 0.5 and 7.5 years \citep{IceCube18}. Extensive studies of TXS 0506+056 have shown that current one-zone
models face significant difficulties in reproducing the neutrino flux of $\sim10^{-11}\:{\rm erg\:cm^{-2}\:s^{-1}}$,
corresponding to the 0.5-year exposure. In contrast, the upper limit associated with the 7.5-year period corresponds
to a more moderate flux of $\sim10^{-12}\:{\rm erg\:cm^{-2}\:s^{-1}}$, which is more consistent with single-zone
model predictions. Therefore, in our modeling, we assume two neutrino data points at the energies corresponding
to the 90\% confidence level lower and upper limits of 183 TeV and 4.3 PeV, respectively, each with a flux level
of $\sim10^{-12}\:{\rm erg\:cm^{-2}\:s^{-1}}$ with an assumed 10\% uncertainty, corresponding to a flat power spectrum.

The modeling of the multimessenger SED of TXS 0506+059 is shown in the left
panel of Figure \ref{fig:sed:txs}. The model, derived using the best-fit parameters, is shown in red, with
its associated uncertainty indicated in gray. The corresponding neutrino spectrum is represented by a red
dashed line. The best-fit parameter values with their uncertainties are provided in the first column of Table \ref{tab:param}.
The electron and proton energy distributions are characterized by power-law indices of $p_e = 1.8$ and
$p_p = 2.0$, respectively. The minimum Lorentz factor of the emitting electrons is $\gamma_{\rm e,min} = 3.3 \times 10^1$.
The fit yields a relatively modest magnetic field strength of $B = 0.03\ \mathrm{G}$,
and a proton maximum Lorentz factor of $\gamma_{\rm p,max} = 2.3 \times 10^6$, favoring a hybrid scenario to explain the
data. The emission region in the jet of TXS 0506+059 is estimated to have a characteristic size of
$R = 5.8 \times 10^{17}\ \mathrm{cm}$, with a Doppler factor $\delta = 15.5$. Regarding the energetics, the proton
luminosity is dominant and very large, estimated to be $L_{\rm p} = 1.5 \times 10^{51}\ \mathrm{erg\ s^{-1}}$, significantly
exceeding the electron luminosity $L_{\rm e} = 3.2 \times 10^{46}\ \mathrm{erg\ s^{-1}}$ and the magnetic field
luminosity $L_{\rm B} = 2.7 \times 10^{44}\ \mathrm{erg\ s^{-1}}$. This indicates that the jet energetics are
predominantly carried by protons, aligning with a hybrid emission scenario where hadronic emission only contributes
indirectly to the electromagnetic spectrum.

The parameters derived in this modeling of TXS 0506+059 are generally consistent
with previous findings in the literature, where a hybrid model was used to explain the multimessenger
emission from this source. For example, in \citet{GBS22}, the SED was modeled using \soprano{} 
and slightly different parameters were obtained, \citep[see Table 1 in][]{GBS22}.
The magnetic field strength obtained here is significantly lower ($B = 0.03$ G) compared to the previous
lepto-hadronic model ($B = 0.57\,{\rm G}$), suggesting weaker magnetization in the emission region. Furthermore,
the Doppler factor ($\delta = 15.5$) and the emission region size ($R = 5.8 \times 10^{17}\ \mathrm{cm}$) indicate
a larger region than previously reported. Despite these differences, the jet energy remains
proton-dominated, consistent with earlier estimates and supporting the hybrid scenario. Although, the scenario
considered and the code used to calculate the SED is the same, the difference is explained by the statistical
approach employed in this paper, which enables the exploration of the parameter space and selection of the best
model. 

This modeling shows that, despite adopting a relatively low neutrino flux
($\sim10^{-12}\:{\rm erg\,cm^{-2}\,s^{-1}}$), the resulting proton luminosity is significantly large,
with $L_{\rm p} = 1.5 \times 10^{51}\ \mathrm{erg\ s^{-1}}$. This result is consistent with previous studies and underscores
the difficulties that one-zone lepto-hadronic models, such as the one used here, face in reproducing a neutrino
flux compatible with a single event detected over approximately 0.5 years of IceCube exposure. Achieving a higher neutrino
flux would require a higher proton luminosities and/or a lower radius, resulting in increased compactness.  
Additionally, the derived proton luminosity exceeds the Eddington luminosity, $\sim4\times10^{46}\:{\rm erg\,s^{-1}}$ for the central black hole in TXS 0506+059 \citep[$3\times10^8\,M_\odot$;][]{2019MNRAS.484L.104P},
posing substantial theoretical difficulties. We further note that the derived proton luminosity
$L_{\rm p} = 1.5 \times 10^{51}\ \mathrm{erg\ s^{-1}}$ is at the upper boundary of the parameter range used
to train our neural network. This high proton luminosity could be mitigated by adopting a two-zone emission scenario
or by including contributions from external photon fields. Indeed, \citet{2019MNRAS.484L.104P} suggested that
TXS 0506+059 is not a typical BL Lac object, but rather a masquerading BL Lac—intrinsically a FSRQ featuring
obscured broad emission lines and a standard accretion disk. Under these conditions, external photon fields can
significantly contribute to the formation of the HE/VHE \gray\ spectrum. Both alternative scenarios will be explored in future works, which will include training a CNN tailored to lepto-hadronic processes that incorporate external photon fields or multi-zones physics.

The modeling presented above requires information about the neutrino spectrum, which is not currently
available from neutrino observations. Therefore, we apply an likelihood that does not
rely on the neutrino spectral shape but instead uses the number of neutrinos detected by IceCube through
the use of the Poisson likelihood, see Section \ref{subsec:likelihood}.
We assumed the observation of one neutrino event over a one-year period and present the results of the fit
in the right panel of Figure \ref{fig:sed:txs}. The corresponding parameters are listed in the middle panel
of Table \ref{tab:param}. To perform this fit, we set $\gamma_{\rm e,min}= 100$. The model based on the best-fit
parameters is shown in red, and the model uncertainty is indicated in gray. The posterior distributions
of the parameters are shown in Appendix in Figure \ref{fig:posterior_TXS}).

Unlike in the case presented above with Gaussian likelihood, the magnetic field
is relatively strong in this scenario, with $B = 5.3 \times 10^2\ \mathrm{G}$,
leading to significant electron cooling by synchrotron radiation. The injected electron distribution is
characterized by a power-law index of $p_{e} = 2.9$ and a moderate cutoff Lorentz factor of
$\gamma_{\rm e,max} = 1.8 \times 10^3$. In this high magnetic field environment, 
synchrotron radiation from protons, injected in an exponentially cut-off power-law with index $p_{p} = 2.0$
and cutoff Lorentz factor of $\gamma_{\rm p,max} = 10^8$ becomes significant and account for the
emission in the HE and VHE $\gamma$-ray bands, leading to a P-sync model. The emission region is relatively compact, with a size of $R = 3.4 \times 10^{14}\ \mathrm{cm}$, and the Doppler factor of the relativistic jet
is $\delta = 23.2$. The jet electron and proton luminosities are
$L_{\rm e} = 9.8 \times 10^{43}\ \mathrm{erg\ s^{-1}}$ and
$L_{\rm p} = 4.9 \times 10^{46}\ \mathrm{erg\ s^{-1}}$ respectively, while
the luminosity associated with the magnetic field is $L_{\rm B} = 6.6 \times 10^{46}\ \mathrm{erg\ s^{-1}}$. The
proton luminosity is significantly lower than the one obtained with the Gaussian likelihood, which lead to a hybrid model. However, in this scenario, the peak of
the neutrino distribution (red dashed line in the right panel of Figure \ref{fig:sed:txs}) is shifted to a higher
energy range of $\sim7\times10^{17}$ eV. The model parameters obtained in this study are generally consistent with those
reported in \citet{GBS22}, for the P-sync model. 

\begin{figure}
    \centering
    \includegraphics[width=0.45\linewidth]{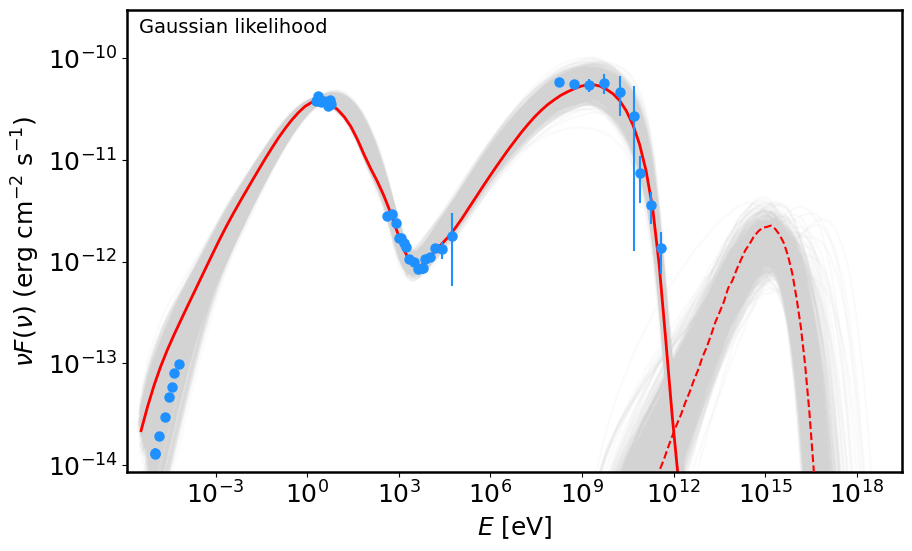}
    \includegraphics[width=0.45\linewidth]{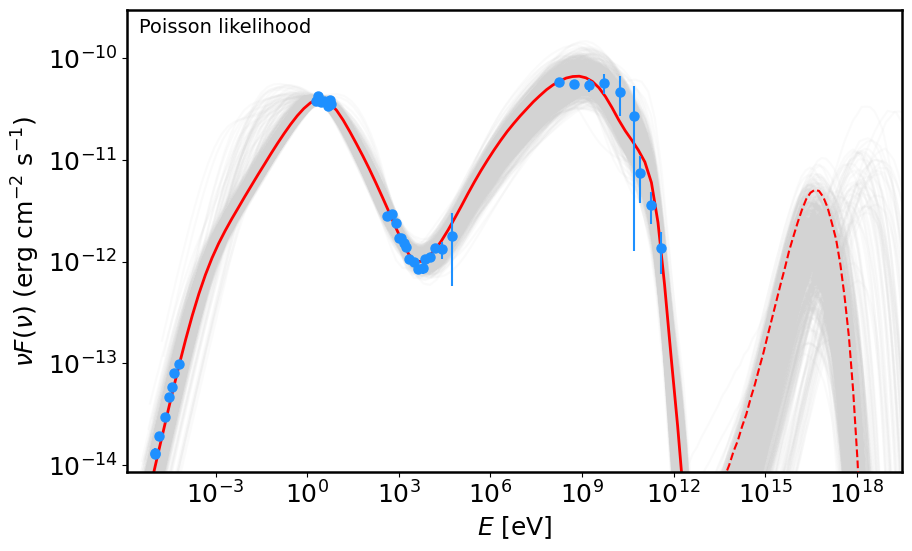}
    \caption{Broadband SED of TXS 0506+059 during the IceCube-170922A neutrino event. The observed data are shown in blue, while the model corresponding to the maximum likelihood parameters is shown in red (dashed red line is the corresponding neutrino spectrum). Model uncertainties are represented in gray, indicating a subset of spectra drawn from the posterior distribution. The model calculations account for EBL absorption using the model of \citet{2011MNRAS.410.2556D}. \textit{Left panel:} SED modeling assuming a predefined neutrino flux level, see text for details. \textit{Right panel:} SED modeling using a Poisson likelihood assuming the detection of one neutrino event by IceCube for a one-year-long exposure.}
    \label{fig:sed:txs}
\end{figure}

\subsection{PKS 0735+178}

The track-like event IceCube-211208A, with an energy of 172 TeV, was detected by IceCube on December 8,
2021 \citep{2021GCN.31191....1I}. Subsequently, additional neutrino events with lower energies were reported
from the same sky region by the Baikal-GVD \citep{Baikal}, the Baksan Underground Scintillation Telescope
\citep{2021ATel15143....1P}, and the KM3NeT undersea neutrino detector \citep{2022ATel15290....1F}. Although the
position of PKS 0735+178 lies slightly outside the error region of IceCube-211208A, a comprehensive multiwavelength
analysis revealed that the source was undergoing significant flaring activity in the optical/UV, X-ray, and \gray\ bands
at the time of the neutrino detections \citep{2023MNRAS.519.1396S}. Given the detection of multiple neutrino events
and the wealth of multiwavelength data collected from various instruments, \citet{2023MNRAS.519.1396S} established
PKS 0735+178 as a strong neutrino-emitting blazar candidate.

We show on Fig. \ref{fig:sed} the SED of PKS 0735+178, retrieved from \citet{2023MNRAS.519.1396S} around the time of the IceCube-211208A event. The best model obtained from the fit is shown in red, while the neutrino spectrum is shown in dashed red.
The fit was performed assuming one neutrino event over a one-year period in the IceCube detector, using a Poisson likelihood approach. The corresponding parameters obtained from the fit are listed in
Table \ref{tab:param}. To reduce the number of free parameters, we fixed $\gamma_{\rm e,min} = 100$. The low-energy
component is explained by synchrotron emission from electrons in a large magnetic field, $B = 7.5 \times 10^2 \ \mathrm{G}$.
In this environment, the electrons are strongly cooled and cannot explain the HE data. The electron injection spectrum
is characterized by a power-law index of $p_{e} = 2.62$ and a maximum energy of $\gamma_{\rm e,max} = 1.26 \times 10^3$.
Instead, the HE data are explained by the radiation from protons,
which have a power-law index of $p_{p} = 1.87$ and a maximum energy of $\gamma_{\rm p,max} = 8.91 \times 10^7$. The modeling indicates
that the emission originates from a compact region with a radius of $R = 3.6 \times 10^{14}\ \mathrm{cm}$ and a Doppler factor
of 26. The fit yields a proton and electron luminosities of
$L_{\rm p} = 8.1 \times 10^{46}\ \mathrm{erg\ s^{-1}}$ and
$L_{\rm e} = 4.1 \times 10^{44}\ \mathrm{erg\ s^{-1}}$ respectively, both lower than the
contribution of the the magnetic field to the jet luminosity,
$L_{\rm B} = 1.8 \times 10^{47}\ \mathrm{erg\ s^{-1}}$. The posterior distributions of the parameters are shown in Appendix in Figure \ref{fig:posterior_PKS0735}.

\begin{figure}
    \centering
    \includegraphics[width=0.5\linewidth]{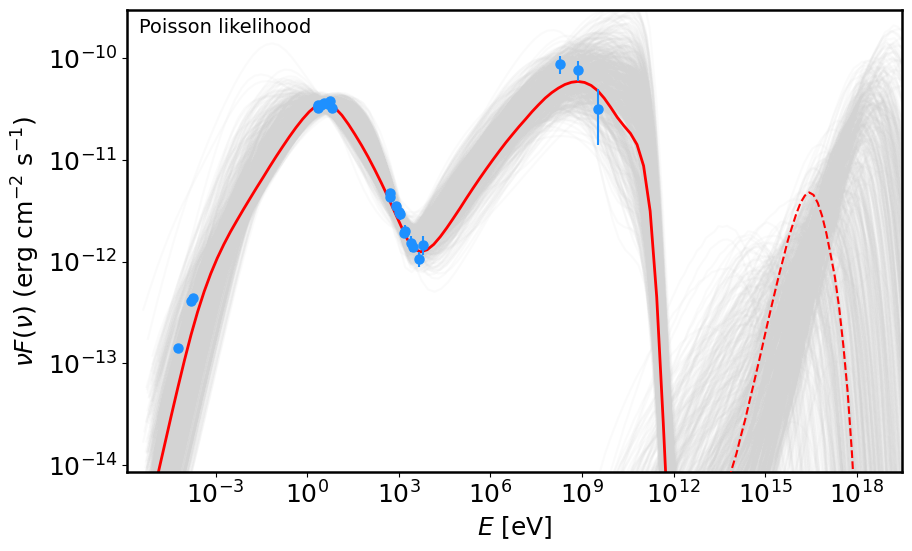}
    \caption{Broadband SED of PKS 0735+178 during the flaring state coincident with the IceCube-211208A event. The fit is performed with a Poisson likelihood for the neutrino and assumes the detection of one neutrino event by IceCube over
    a one-year-long exposure. The color coding is the same as in Figure \ref{fig:sed:txs}.}
    \label{fig:sed}
\end{figure}

The parameters derived from the current fit are generally consistent with those reported in \citet{2023MNRAS.519.1396S}, who investigated
three different scenarios for modeling the multimessenger SED of PKS 0735+178, including a P-syn
model that is broadly similar to the one obtained here. There are notable differences in the electron distributions, primarily
due to the treatment of the cut-off and maximum energies. Indeed, in contrast with \citep{2023MNRAS.519.1396S}, we do not consider a maximum electron Lorentz factor, such that $Q_e(\gamma_e) = 0$ if $\gamma_e > \gamma_{e,\rm c}$ for the electron
injection function.
This distinction affects the shape of the high-energy tail and leads to differences in the overall electron distribution.
In contrast, the characteristics of the proton distribution obtained in this fit,
$p_p = 1.87$ and $\gamma_{\rm p,max} \approx 10^8$, match the parameters adopted in the previous study, where a spectral
index of 2.0 and a maximum Lorentz factor of $3.0 \times 10^8$ were used. The fit also yielded a smaller emitting region
size, $R \approx 3.55 \times 10^{14}\:{\rm cm}$, compared to the value used in \citet{2023MNRAS.519.1396S} ($2.8 \times 10^{15}\:{\rm cm}$),
leading to differences in the derived magnetic field strength and Doppler factor. Despite these differences in individual model
parameters, the resulting electron and proton luminosities show good overall agreement with those reported in the earlier
study, supporting the consistency of the physical interpretation across different modeling approaches.

Although the best-fit parameters are consistent with the P-syn model, the posterior distributions reveal a bimodal
structure. The secondary, less probable mode corresponds to parameter values characteristic of the hybrid model.
This bimodality can only be explored through a complete sampling of the posterior. 
In the present analysis, the data are not sufficiently constraining to clearly distinguish between the P-syn and hybrid models.
Improved data treatment, particularly at the highest photon energies, or alternatively mode data may help resolve this ambiguity and better separate the two scenarios.

\section{Hadronic modeling available in \texttt{MMDC}}\label{sec:aval}

Considering the importance of making advanced multimessenger modeling tools accessible to the broader
scientific community, the CNN developed for hadronic and lepto-hadronic modeling is  available through the
\texttt{MMDC} platform \citep[www.mmdc.am;][]{2024AJ....168..289S}, alongside previously released CNNs for SSC
\citep{BSD23} and EIC \citep{SBC24} scenarios. \texttt{MMDC} is a novel web-based platform designed to support
the retrieval, visualization, and theoretical interpretation of multiwavelength and multimessenger data from
blazar. It integrates extensive archival and newly processed data and provides tools for
constructing and analyzing time-resolved SEDs.

The CNN for hadronic scenarios enables users to perform fast and
self-consistent modeling of blazar SEDs, including electromagnetic and neutrino emissions. Users can upload their
own data in the specified format or retrieve them directly from \texttt{MMDC} and choose between two
fitting approaches: \textit{(i)} providing a neutrino spectrum at selected energies, or \textit{(ii)} applying a
Poisson likelihood based on the expected number of neutrino events over a specified time spam. Once submitted,
the fit is carried out on dedicated compute resources, and the results—including best-fit 
parameters, model predictions, and posterior distributions—are returned to the user via email and accessible
through the \texttt{MMDC} interface.

By offering this tool through \texttt{MMDC}, the framework empowers the community to perform in-depth modeling
of multimessenger data under both leptonic and hadronic scenarios. This capability is particularly valuable in
this data-rich era and is expected to significantly contribute to the interpretation of blazar emission and jet physics.

\section{Conclusion}\label{sec:conc}

The era of multimessenger observations is transforming how we study astrophysical sources. By combining data from
electromagnetic radiation and neutrinos, it is now possible to
gain a deeper understanding of extreme cosmic environments, such as the jets of blazars. However, from a numerical point of view, the current codes—although robust and physically complete for
modeling lepto-hadronic processes—face significant computational challenges. In particular, these models are
time-consuming to compute, limiting their applicability for parameter space exploration and data fitting.

To overcome these limitations, and following the methodologies of \citet{BSD23} and \citet{SBC24}, we developed a
CNN trained on synthetic spectra generated using \soprano{}, which self-consistently
solves the time-dependent kinetic equations for particles and photons. \soprano{} incorporates all relevant leptonic
and hadronic processes, including synchrotron radiation, inverse Compton scattering, Bethe–Heitler pair production,
and photo-pion production, simulating both electromagnetic and neutrino emissions. By training the CNN on outputs
generated from a broad, physically motivated parameter space, the network captures the behavior of both primary
and secondary particles and can be used to model multimessenger data in hadronic setups, including both P-syn
and hybrid scenarios.

The trained CNN offers a fast and efficient alternative to traditional numerical modeling. Once trained, it is coupled
with Bayesian inference tools (e.g., MultiNest) to fit multimessenger data and derive physical parameters of blazar jets,
taking into account constraints from both electromagnetic and neutrino observations. As a demonstration, we applied the
trained network to the SEDs of TXS 0506+059 and PKS 0735+178, two of the most promising neutrino-emitting blazars.
For TXS 0506+059, two fitting approaches were used: one assuming a flat $E_{\nu}^{-2}$-type neutrino spectrum across
two energy bands, which results favor a hybrid model; and another using a Poisson likelihood based
on the number of detected neutrinos, which favored a proton synchrotron scenario. This demonstrates (i) the versatility
of the developed method in comparing and testing both P-syn and hybrid models, and in retrieving
the corresponding best-fit parameters (ii) the need for a better data treatment at HE part of the electromagnetic spectrum, as well as for neutrinos, and (iii) the potential of using multimessenger data to constrain the properties of relativistic jets.

The CNN provides an efficient and reusable solution for modeling multimessenger blazar data. As the network
is publicly available through the \texttt{MMDC} platform, along with CNNs for the SSC \citep{BSD23} and EIC models
\citep{SBC24}, it can be widely used by the community. This is particularly important in light of upcoming
observatories such as the CTAO, KM3NeT, and IceCube-Gen2, which are expected to
significantly expand the available multimessenger dataset.

A current limitation of the trained CNN for the hadronic model is that it does not yet include
the effects of external photon fields. These fields can influence the cooling of electrons and protons and are
especially relevant for sources with standard accretion disks or broad-line regions (e.g., FSRQs). As a next step,
we plan to create a new model which will
include external radiation fields, enabling the CNN to capture an even broader range of physical conditions
relevant to blazar emission.

In summary, the novel method proposed by \citet{BSD23}, and later extended to include external inverse Compton
scenarios in \citet{SBC24}, is further advanced in this work by incorporating a significantly more complex
hadronic model. The number of physical processes and particle species involved in
the hadronic case far exceeds those in leptonic cases, yet the developed approach
still yields a robust and reliable neural network capable of comprehensive modeling. The resulting networks enable
state-of-the-art, self-consistent analyses of multi-wavelength and multimessenger data from blazar observations,
offering a powerful tool for interpreting the growing datasets provided by current and upcoming observatories.

\section*{Acknowledgements}
NS, VV and MK acknowledge the support by the Higher Education and Science
Committee of the Republic of Armenia, in the frames of the
research project No 23LCG-1C004.

\section*{Data availability}
All the observational data used in this paper is public. The convolutional neural network used
to fit the SEDs can be shared on a reasonable request to the corresponding author. In addition, it is publicly available through the Markarian Multiwavelength
Data Center (\url{http://www.mmdc.am}).

\bibliographystyle{mnras}
\bibliography{biblio}

\begin{thebibliography}{}
\makeatletter
\relax
\def\mn@urlcharsother{\let\do\@makeother \do\$\do\&\do\#\do\^\do\_\do\%\do\~}
\def\mn@doi{\begingroup\mn@urlcharsother \@ifnextchar [ {\mn@doi@}
  {\mn@doi@[]}}
\def\mn@doi@[#1]#2{\def\@tempa{#1}\ifx\@tempa\@empty \href
  {http://dx.doi.org/#2} {doi:#2}\else \href {http://dx.doi.org/#2} {#1}\fi
  \endgroup}
\def\mn@eprint#1#2{\mn@eprint@#1:#2::\@nil}
\def\mn@eprint@arXiv#1{\href {http://arxiv.org/abs/#1} {{\tt arXiv:#1}}}
\def\mn@eprint@dblp#1{\href {http://dblp.uni-trier.de/rec/bibtex/#1.xml}
  {dblp:#1}}
\def\mn@eprint@#1:#2:#3:#4\@nil{\def\@tempa {#1}\def\@tempb {#2}\def\@tempc
  {#3}\ifx \@tempc \@empty \let \@tempc \@tempb \let \@tempb \@tempa \fi \ifx
  \@tempb \@empty \def\@tempb {arXiv}\fi \@ifundefined
  {mn@eprint@\@tempb}{\@tempb:\@tempc}{\expandafter \expandafter \csname
  mn@eprint@\@tempb\endcsname \expandafter{\@tempc}}}

\bibitem[\protect\citeauthoryear{{Abe} et~al.,}{{Abe}
  et~al.}{2023}]{2023ApJS..266...37A}
{Abe} H.,  et~al., 2023, \mn@doi [\apjs] {10.3847/1538-4365/acc181}, \href
  {https://ui.adsabs.harvard.edu/abs/2023ApJS..266...37A} {266, 37}

\bibitem[\protect\citeauthoryear{{Acharyya} et~al.,}{{Acharyya}
  et~al.}{2023}]{2023ApJ...954...70A}
{Acharyya} A.,  et~al., 2023, \mn@doi [\apj] {10.3847/1538-4357/ace327}, \href
  {https://ui.adsabs.harvard.edu/abs/2023ApJ...954...70A} {954, 70}

\bibitem[\protect\citeauthoryear{{Ansoldi} et~al.,}{{Ansoldi}
  et~al.}{2018}]{2018ApJ...863L..10A}
{Ansoldi} S.,  et~al., 2018, \mn@doi [\apjl] {10.3847/2041-8213/aad083}, \href
  {http://adsabs.harvard.edu/abs/2018ApJ...863L..10A} {863, L10}

\bibitem[\protect\citeauthoryear{{Becker}}{{Becker}}{2008}]{2008PhR...458..173B}
{Becker} J.~K.,  2008, \mn@doi [\physrep] {10.1016/j.physrep.2007.10.006},
  \href {https://ui.adsabs.harvard.edu/abs/2008PhR...458..173B} {458, 173}

\bibitem[\protect\citeauthoryear{{B{\'e}gu{\'e}}, {Sahakyan},
  {Dereli-B{\'e}gu{\'e}}, {Giommi}, {Gasparyan}, {Khachatryan}, {Casotto}  \&
  {Pe'er}}{{B{\'e}gu{\'e}} et~al.}{2024}]{BSD23}
{B{\'e}gu{\'e}} D.,  {Sahakyan} N.,  {Dereli-B{\'e}gu{\'e}} H.,  {Giommi} P.,
  {Gasparyan} S.,  {Khachatryan} M.,  {Casotto} A.,   {Pe'er} A.,  2024,
  \mn@doi [\apj] {10.3847/1538-4357/ad19cf}, \href
  {https://ui.adsabs.harvard.edu/abs/2024ApJ...963...71B} {963, 71}

\bibitem[\protect\citeauthoryear{{Blaufuss}, {Kintscher}, {Lu}  \&
  {Tung}}{{Blaufuss} et~al.}{2019}]{2019ICRC...36.1021B}
{Blaufuss} E.,  {Kintscher} T.,  {Lu} L.,   {Tung} C.~F.,  2019, in 36th
  International Cosmic Ray Conference (ICRC2019). p.~1021 (\mn@eprint {arXiv}
  {1908.04884}), \mn@doi{10.22323/1.358.01021}

\bibitem[\protect\citeauthoryear{{B{\l}a{\.z}ejowski}, {Sikora}, {Moderski}  \&
  {Madejski}}{{B{\l}a{\.z}ejowski} et~al.}{2000}]{2000ApJ...545..107B}
{B{\l}a{\.z}ejowski} M.,  {Sikora} M.,  {Moderski} R.,   {Madejski} G.~M.,
  2000, \mn@doi [\apj] {10.1086/317791}, \href
  {https://ui.adsabs.harvard.edu/abs/2000ApJ...545..107B} {545, 107}

\bibitem[\protect\citeauthoryear{{Bloom} \& {Marscher}}{{Bloom} \&
  {Marscher}}{1996}]{1996ApJ...461..657B}
{Bloom} S.~D.,  {Marscher} A.~P.,  1996, \mn@doi [\apj] {10.1086/177092}, \href
  {https://ui.adsabs.harvard.edu/abs/1996ApJ...461..657B} {461, 657}

\bibitem[\protect\citeauthoryear{{Boersma} \& {van Leeuwen}}{{Boersma} \& {van
  Leeuwen}}{2023}]{BvL23}
{Boersma} O.~M.,  {van Leeuwen} J.,  2023, \mn@doi [\pasa]
  {10.1017/pasa.2023.32}, \href
  {https://ui.adsabs.harvard.edu/abs/2023PASA...40...30B} {40, e030}

\bibitem[\protect\citeauthoryear{{B{\"o}ttcher}, {Reimer}, {Sweeney}  \&
  {Prakash}}{{B{\"o}ttcher} et~al.}{2013}]{2013ApJ...768...54B}
{B{\"o}ttcher} M.,  {Reimer} A.,  {Sweeney} K.,   {Prakash} A.,  2013, \mn@doi
  [\apj] {10.1088/0004-637X/768/1/54}, \href
  {http://adsabs.harvard.edu/abs/2013ApJ...768...54B} {768, 54}

\bibitem[\protect\citeauthoryear{{Burgess}}{{Burgess}}{2023}]{Bur23}
{Burgess} J.~M.,  2023, \mn@doi [The Journal of Open Source Software]
  {10.21105/joss.04969}, \href
  {https://ui.adsabs.harvard.edu/abs/2023JOSS....8.4969B} {8, 4969}

\bibitem[\protect\citeauthoryear{{Cash}}{{Cash}}{1979}]{Cash79}
{Cash} W.,  1979, \mn@doi [\apj] {10.1086/156922}, \href
  {https://ui.adsabs.harvard.edu/abs/1979ApJ...228..939C} {228, 939}

\bibitem[\protect\citeauthoryear{{Cerruti}, {Zech}, {Boisson}, {Emery}, {Inoue}
   \& {Lenain}}{{Cerruti} et~al.}{2019}]{2019MNRAS.483L..12C}
{Cerruti} M.,  {Zech} A.,  {Boisson} C.,  {Emery} G.,  {Inoue} S.,   {Lenain}
  J.-P.,  2019, \mn@doi [\mnras] {10.1093/mnrasl/sly210}, \href
  {https://ui.adsabs.harvard.edu/abs/2019MNRAS.483L..12C} {483, L12}

\bibitem[\protect\citeauthoryear{{Dermer} \& {Schlickeiser}}{{Dermer} \&
  {Schlickeiser}}{1994}]{1994ApJS...90..945D}
{Dermer} C.~D.,  {Schlickeiser} R.,  1994, \mn@doi [\apjs] {10.1086/191929},
  \href {https://ui.adsabs.harvard.edu/abs/1994ApJS...90..945D} {90, 945}

\bibitem[\protect\citeauthoryear{{Dermer}, {Schlickeiser}  \&
  {Mastichiadis}}{{Dermer} et~al.}{1992}]{1992A&A...256L..27D}
{Dermer} C.~D.,  {Schlickeiser} R.,   {Mastichiadis} A.,  1992, \aap, \href
  {https://ui.adsabs.harvard.edu/abs/1992A&A...256L..27D} {256, L27}

\bibitem[\protect\citeauthoryear{{Dom{\'\i}nguez} et~al.,}{{Dom{\'\i}nguez}
  et~al.}{2011}]{2011MNRAS.410.2556D}
{Dom{\'\i}nguez} A.,  et~al., 2011, \mn@doi [\mnras]
  {10.1111/j.1365-2966.2010.17631.x}, \href
  {https://ui.adsabs.harvard.edu/abs/2011MNRAS.410.2556D} {410, 2556}

\bibitem[\protect\citeauthoryear{{Dzhilkibaev}, {Suvorova}  \& {Baikal-GVD
  Collaboration}}{{Dzhilkibaev} et~al.}{2021}]{Baikal}
{Dzhilkibaev} Z.~A.,  {Suvorova} O.,   {Baikal-GVD Collaboration} 2021, The
  Astronomer's Telegram, \href
  {https://ui.adsabs.harvard.edu/abs/2021ATel15112....1D} {15112, 1}

\bibitem[\protect\citeauthoryear{{Fantini}, {Gallo Rosso}, {Vissani}  \&
  {Zema}}{{Fantini} et~al.}{2018}]{FGV18}
{Fantini} G.,  {Gallo Rosso} A.,  {Vissani} F.,   {Zema} V.,  2018, \mn@doi
  [arXiv e-prints] {10.48550/arXiv.1802.05781}, \href
  {https://ui.adsabs.harvard.edu/abs/2018arXiv180205781F} {p. arXiv:1802.05781}

\bibitem[\protect\citeauthoryear{{Feroz}, {Hobson}  \& {Bridges}}{{Feroz}
  et~al.}{2009}]{FHB09}
{Feroz} F.,  {Hobson} M.~P.,   {Bridges} M.,  2009, \mn@doi [\mnras]
  {10.1111/j.1365-2966.2009.14548.x}, \href
  {http://adsabs.harvard.edu/abs/2009MNRAS.398.1601F} {398, 1601}

\bibitem[\protect\citeauthoryear{{Filippini} et~al.,}{{Filippini}
  et~al.}{2022}]{2022ATel15290....1F}
{Filippini} F.,  et~al., 2022, The Astronomer's Telegram, \href
  {https://ui.adsabs.harvard.edu/abs/2022ATel15290....1F} {15290, 1}

\bibitem[\protect\citeauthoryear{{Gao}, {Fedynitch}, {Winter}  \& {Pohl}}{{Gao}
  et~al.}{2019}]{2019NatAs...3...88G}
{Gao} S.,  {Fedynitch} A.,  {Winter} W.,   {Pohl} M.,  2019, \mn@doi [Nature
  Astronomy] {10.1038/s41550-018-0610-1}, \href
  {https://ui.adsabs.harvard.edu/abs/2019NatAs...3...88G} {3, 88}

\bibitem[\protect\citeauthoryear{{Gasparyan}, {B{\'e}gu{\'e}}  \&
  {Sahakyan}}{{Gasparyan} et~al.}{2022}]{GBS22}
{Gasparyan} S.,  {B{\'e}gu{\'e}} D.,   {Sahakyan} N.,  2022, \mn@doi [\mnras]
  {10.1093/mnras/stab2688}, \href
  {https://ui.adsabs.harvard.edu/abs/2022MNRAS.509.2102G} {509, 2102}

\bibitem[\protect\citeauthoryear{{Ghisellini}, {Maraschi}  \&
  {Treves}}{{Ghisellini} et~al.}{1985}]{1985A&A...146..204G}
{Ghisellini} G.,  {Maraschi} L.,   {Treves} A.,  1985, \aap, \href
  {https://ui.adsabs.harvard.edu/abs/1985A&A...146..204G} {146, 204}

\bibitem[\protect\citeauthoryear{{Giommi}, {Padovani}, {Oikonomou}, {Glauch},
  {Paiano}  \& {Resconi}}{{Giommi} et~al.}{2020}]{2020A&A...640L...4G}
{Giommi} P.,  {Padovani} P.,  {Oikonomou} F.,  {Glauch} T.,  {Paiano} S.,
  {Resconi} E.,  2020, \mn@doi [\aap] {10.1051/0004-6361/202038423}, \href
  {https://ui.adsabs.harvard.edu/abs/2020A&A...640L...4G} {640, L4}

\bibitem[\protect\citeauthoryear{{IceCube Collaboration}}{{IceCube
  Collaboration}}{2021}]{2021GCN.31191....1I}
{IceCube Collaboration} 2021, GRB Coordinates Network, \href
  {https://ui.adsabs.harvard.edu/abs/2021GCN.31191....1I} {31191, 1}

\bibitem[\protect\citeauthoryear{{IceCube Collaboration} et~al.,}{{IceCube
  Collaboration} et~al.}{2018a}]{IceCube18}
{IceCube Collaboration} et~al., 2018a, \mn@doi [Science]
  {10.1126/science.aat1378}, \href
  {http://adsabs.harvard.edu/abs/2018Sci...361.1378I} {361, eaat1378}

\bibitem[\protect\citeauthoryear{{IceCube Collaboration} et~al.,}{{IceCube
  Collaboration} et~al.}{2018b}]{IceCube18b}
{IceCube Collaboration} et~al., 2018b, \mn@doi [Science]
  {10.1126/science.aat2890}, \href
  {https://ui.adsabs.harvard.edu/abs/2018Sci...361..147I} {361, 147}

\bibitem[\protect\citeauthoryear{{Keivani} et~al.,}{{Keivani}
  et~al.}{2018}]{2018ApJ...864...84K}
{Keivani} A.,  et~al., 2018, \mn@doi [\apj] {10.3847/1538-4357/aad59a}, \href
  {http://adsabs.harvard.edu/abs/2018ApJ...864...84K} {864, 84}

\bibitem[\protect\citeauthoryear{{Krau{\ss}} et~al.,}{{Krau{\ss}}
  et~al.}{2020}]{2020MNRAS.497.2553K}
{Krau{\ss}} F.,  et~al., 2020, \mn@doi [\mnras] {10.1093/mnras/staa2148}, \href
  {https://ui.adsabs.harvard.edu/abs/2020MNRAS.497.2553K} {497, 2553}

\bibitem[\protect\citeauthoryear{{Liao} et~al.,}{{Liao}
  et~al.}{2022}]{2022ApJ...932L..25L}
{Liao} N.-H.,  et~al., 2022, \mn@doi [\apjl] {10.3847/2041-8213/ac756f}, \href
  {https://ui.adsabs.harvard.edu/abs/2022ApJ...932L..25L} {932, L25}

\bibitem[\protect\citeauthoryear{{Mannheim}}{{Mannheim}}{1993}]{1993A&A...269...67M}
{Mannheim} K.,  1993, \aap, \href
  {http://adsabs.harvard.edu/abs/1993A%26A...269...67M} {269, 67}

\bibitem[\protect\citeauthoryear{{Mannheim} \& {Biermann}}{{Mannheim} \&
  {Biermann}}{1989}]{1989A&A...221..211M}
{Mannheim} K.,  {Biermann} P.~L.,  1989, \aap, \href
  {http://adsabs.harvard.edu/abs/1989A%26A...221..211M} {221, 211}

\bibitem[\protect\citeauthoryear{{Maraschi}, {Ghisellini}  \&
  {Celotti}}{{Maraschi} et~al.}{1992}]{1992ApJ...397L...5M}
{Maraschi} L.,  {Ghisellini} G.,   {Celotti} A.,  1992, \mn@doi [\apjl]
  {10.1086/186531}, \href
  {https://ui.adsabs.harvard.edu/abs/1992ApJ...397L...5M} {397, L5}

\bibitem[\protect\citeauthoryear{McKay, Beckman  \& Conover}{McKay
  et~al.}{2000}]{MBC00}
McKay M.~D.,  Beckman R.~J.,   Conover W.~J.,  2000, Technometrics, 42, 55

\bibitem[\protect\citeauthoryear{{M{\"u}cke} \& {Protheroe}}{{M{\"u}cke} \&
  {Protheroe}}{2001}]{2001APh....15..121M}
{M{\"u}cke} A.,  {Protheroe} R.~J.,  2001, \mn@doi [Astroparticle Physics]
  {10.1016/S0927-6505(00)00141-9}, \href
  {https://ui.adsabs.harvard.edu/abs/2001APh....15..121M} {15, 121}

\bibitem[\protect\citeauthoryear{{M{\"u}cke}, {Protheroe}, {Engel}, {Rachen}
  \& {Stanev}}{{M{\"u}cke} et~al.}{2003}]{mucke2}
{M{\"u}cke} A.,  {Protheroe} R.~J.,  {Engel} R.,  {Rachen} J.~P.,   {Stanev}
  T.,  2003, \mn@doi [Astroparticle Physics] {10.1016/S0927-6505(02)00185-8},
  \href {http://adsabs.harvard.edu/abs/2003APh....18..593M} {18, 593}

\bibitem[\protect\citeauthoryear{{Murase}, {Oikonomou}  \&
  {Petropoulou}}{{Murase} et~al.}{2018}]{2018ApJ...865..124M}
{Murase} K.,  {Oikonomou} F.,   {Petropoulou} M.,  2018, \mn@doi [\apj]
  {10.3847/1538-4357/aada00}, \href
  {http://adsabs.harvard.edu/abs/2018ApJ...865..124M} {865, 124}

\bibitem[\protect\citeauthoryear{{Oikonomou}, {Petropoulou}, {Murase},
  {Tohuvavohu}, {Vasilopoulos}, {Buson}  \& {Santander}}{{Oikonomou}
  et~al.}{2021}]{2021JCAP...10..082O}
{Oikonomou} F.,  {Petropoulou} M.,  {Murase} K.,  {Tohuvavohu} A.,
  {Vasilopoulos} G.,  {Buson} S.,   {Santander} M.,  2021, \mn@doi [\jcap]
  {10.1088/1475-7516/2021/10/082}, \href
  {https://ui.adsabs.harvard.edu/abs/2021JCAP...10..082O} {2021, 082}

\bibitem[\protect\citeauthoryear{{Omeliukh} et~al.,}{{Omeliukh}
  et~al.}{2025}]{2025A&A...695A.266O}
{Omeliukh} A.,  et~al., 2025, \mn@doi [\aap] {10.1051/0004-6361/202452143},
  \href {https://ui.adsabs.harvard.edu/abs/2025A&A...695A.266O} {695, A266}

\bibitem[\protect\citeauthoryear{{Padovani}, {Giommi}, {Resconi}, {Glauch},
  {Arsioli}, {Sahakyan}  \& {Huber}}{{Padovani}
  et~al.}{2018}]{2018MNRAS.480..192P}
{Padovani} P.,  {Giommi} P.,  {Resconi} E.,  {Glauch} T.,  {Arsioli} B.,
  {Sahakyan} N.,   {Huber} M.,  2018, \mn@doi [\mnras] {10.1093/mnras/sty1852},
  \href {https://ui.adsabs.harvard.edu/abs/2018MNRAS.480..192P} {480, 192}

\bibitem[\protect\citeauthoryear{{Padovani}, {Oikonomou}, {Petropoulou},
  {Giommi}  \& {Resconi}}{{Padovani} et~al.}{2019}]{2019MNRAS.484L.104P}
{Padovani} P.,  {Oikonomou} F.,  {Petropoulou} M.,  {Giommi} P.,   {Resconi}
  E.,  2019, \mn@doi [\mnras] {10.1093/mnrasl/slz011}, \href
  {https://ui.adsabs.harvard.edu/abs/2019MNRAS.484L.104P} {484, L104}

\bibitem[\protect\citeauthoryear{{Padovani}, {Boccardi}, {Falomo}  \&
  {Giommi}}{{Padovani} et~al.}{2022}]{2022MNRAS.511.4697P}
{Padovani} P.,  {Boccardi} B.,  {Falomo} R.,   {Giommi} P.,  2022, \mn@doi
  [\mnras] {10.1093/mnras/stac376}, \href
  {https://ui.adsabs.harvard.edu/abs/2022MNRAS.511.4697P} {511, 4697}

\bibitem[\protect\citeauthoryear{{Paiano}, {Falomo}, {Treves}  \&
  {Scarpa}}{{Paiano} et~al.}{2018}]{2018ApJ...854L..32P}
{Paiano} S.,  {Falomo} R.,  {Treves} A.,   {Scarpa} R.,  2018, \mn@doi [\apjl]
  {10.3847/2041-8213/aaad5e}, \href
  {https://ui.adsabs.harvard.edu/abs/2018ApJ...854L..32P} {854, L32}

\bibitem[\protect\citeauthoryear{{Paliya}, {B{\"o}ttcher}, {Olmo-Garc{\'\i}a},
  {Dom{\'\i}nguez}, {Gil de Paz}, {Franckowiak}, {Garrappa}  \&
  {Stein}}{{Paliya} et~al.}{2020}]{2020ApJ...902...29P}
{Paliya} V.~S.,  {B{\"o}ttcher} M.,  {Olmo-Garc{\'\i}a} A.,  {Dom{\'\i}nguez}
  A.,  {Gil de Paz} A.,  {Franckowiak} A.,  {Garrappa} S.,   {Stein} R.,  2020,
  \mn@doi [\apj] {10.3847/1538-4357/abb46e}, \href
  {https://ui.adsabs.harvard.edu/abs/2020ApJ...902...29P} {902, 29}

\bibitem[\protect\citeauthoryear{{Petkov}, {Novoseltsev}, {Novoseltseva}  \&
  {Baksan Underground Scintillation Telescope Group}}{{Petkov}
  et~al.}{2021}]{2021ATel15143....1P}
{Petkov} V.~B.,  {Novoseltsev} Y.~F.,  {Novoseltseva} R.~V.,   {Baksan
  Underground Scintillation Telescope Group} 2021, The Astronomer's Telegram,
  \href {https://ui.adsabs.harvard.edu/abs/2021ATel15143....1P} {15143, 1}

\bibitem[\protect\citeauthoryear{{Petropoulou} \& {Mastichiadis}}{{Petropoulou}
  \& {Mastichiadis}}{2015}]{2015MNRAS.447...36P}
{Petropoulou} M.,  {Mastichiadis} A.,  2015, \mn@doi [\mnras]
  {10.1093/mnras/stu2364}, \href
  {https://ui.adsabs.harvard.edu/abs/2015MNRAS.447...36P} {447, 36}

\bibitem[\protect\citeauthoryear{{Petropoulou}, {Oikonomou}, {Mastichiadis},
  {Murase}, {Padovani}, {Vasilopoulos}  \& {Giommi}}{{Petropoulou}
  et~al.}{2020}]{2020ApJ...899..113P}
{Petropoulou} M.,  {Oikonomou} F.,  {Mastichiadis} A.,  {Murase} K.,
  {Padovani} P.,  {Vasilopoulos} G.,   {Giommi} P.,  2020, \mn@doi [\apj]
  {10.3847/1538-4357/aba8a0}, \href
  {https://ui.adsabs.harvard.edu/abs/2020ApJ...899..113P} {899, 113}

\bibitem[\protect\citeauthoryear{{Prince}, {Das}, {Gupta}, {Majumdar}  \&
  {Czerny}}{{Prince} et~al.}{2024}]{2024MNRAS.527.8746P}
{Prince} R.,  {Das} S.,  {Gupta} N.,  {Majumdar} P.,   {Czerny} B.,  2024,
  \mn@doi [\mnras] {10.1093/mnras/stad3804}, \href
  {https://ui.adsabs.harvard.edu/abs/2024MNRAS.527.8746P} {527, 8746}

\bibitem[\protect\citeauthoryear{{Righi}, {Tavecchio}  \& {Pacciani}}{{Righi}
  et~al.}{2019}]{2019MNRAS.484.2067R}
{Righi} C.,  {Tavecchio} F.,   {Pacciani} L.,  2019, \mn@doi [\mnras]
  {10.1093/mnras/sty3072}, \href
  {https://ui.adsabs.harvard.edu/abs/2019MNRAS.484.2067R} {484, 2067}

\bibitem[\protect\citeauthoryear{{Rodrigues}, {Paliya}, {Garrappa}, {Omeliukh},
  {Franckowiak}  \& {Winter}}{{Rodrigues} et~al.}{2024a}]{RPG24}
{Rodrigues} X.,  {Paliya} V.~S.,  {Garrappa} S.,  {Omeliukh} A.,  {Franckowiak}
  A.,   {Winter} W.,  2024a, \mn@doi [\aap] {10.1051/0004-6361/202347540},
  \href {https://ui.adsabs.harvard.edu/abs/2024A\&A...681A.119R} {681, A119}

\bibitem[\protect\citeauthoryear{{Rodrigues}, {Karl}, {Padovani}, {Giommi},
  {Paiano}, {Falomo}, {Petropoulou}  \& {Oikonomou}}{{Rodrigues}
  et~al.}{2024b}]{RKP24}
{Rodrigues} X.,  {Karl} M.,  {Padovani} P.,  {Giommi} P.,  {Paiano} S.,
  {Falomo} R.,  {Petropoulou} M.,   {Oikonomou} F.,  2024b, \mn@doi [\aap]
  {10.1051/0004-6361/202450592}, \href
  {https://ui.adsabs.harvard.edu/abs/2024A\&A...689A.147R} {689, A147}

\bibitem[\protect\citeauthoryear{{Sahakyan}}{{Sahakyan}}{2018}]{2018ApJ...866..109S}
{Sahakyan} N.,  2018, \mn@doi [\apj] {10.3847/1538-4357/aadade}, \href
  {http://adsabs.harvard.edu/abs/2018ApJ...866..109S} {866, 109}

\bibitem[\protect\citeauthoryear{{Sahakyan}}{{Sahakyan}}{2019}]{2019A&A...622A.144S}
{Sahakyan} N.,  2019, \mn@doi [\aap] {10.1051/0004-6361/201834606}, \href
  {https://ui.adsabs.harvard.edu/abs/2019A&A...622A.144S} {622, A144}

\bibitem[\protect\citeauthoryear{{Sahakyan}, {Giommi}, {Padovani},
  {Petropoulou}, {B{\'e}gu{\'e}}, {Boccardi}  \& {Gasparyan}}{{Sahakyan}
  et~al.}{2023}]{2023MNRAS.519.1396S}
{Sahakyan} N.,  {Giommi} P.,  {Padovani} P.,  {Petropoulou} M.,
  {B{\'e}gu{\'e}} D.,  {Boccardi} B.,   {Gasparyan} S.,  2023, \mn@doi [\mnras]
  {10.1093/mnras/stac3607}, \href
  {https://ui.adsabs.harvard.edu/abs/2023MNRAS.519.1396S} {519, 1396}

\bibitem[\protect\citeauthoryear{{Sahakyan} et~al.,}{{Sahakyan}
  et~al.}{2024a}]{2024AJ....168..289S}
{Sahakyan} N.,  et~al., 2024a, \mn@doi [\aj] {10.3847/1538-3881/ad8231}, \href
  {https://ui.adsabs.harvard.edu/abs/2024AJ....168..289S} {168, 289}

\bibitem[\protect\citeauthoryear{{Sahakyan} et~al.,}{{Sahakyan}
  et~al.}{2024b}]{SBC24}
{Sahakyan} N.,  et~al., 2024b, \mn@doi [\apj] {10.3847/1538-4357/ad5351}, \href
  {https://ui.adsabs.harvard.edu/abs/2024ApJ...971...70S} {971, 70}

\bibitem[\protect\citeauthoryear{{Sikora}, {Begelman}  \& {Rees}}{{Sikora}
  et~al.}{1994}]{1994ApJ...421..153S}
{Sikora} M.,  {Begelman} M.~C.,   {Rees} M.~J.,  1994, \mn@doi [\apj]
  {10.1086/173633}, \href
  {https://ui.adsabs.harvard.edu/abs/1994ApJ...421..153S} {421, 153}

\bibitem[\protect\citeauthoryear{{Tzavellas}, {Vasilopoulos}, {Petropoulou},
  {Mastichiadis}  \& {Stathopoulos}}{{Tzavellas} et~al.}{2024}]{TVP24}
{Tzavellas} A.,  {Vasilopoulos} G.,  {Petropoulou} M.,  {Mastichiadis} A.,
  {Stathopoulos} S.~I.,  2024, \mn@doi [\aap] {10.1051/0004-6361/202348566},
  \href {https://ui.adsabs.harvard.edu/abs/2024A&A...683A.185T} {683, A185}

\bibitem[\protect\citeauthoryear{{Urry} \& {Padovani}}{{Urry} \&
  {Padovani}}{1995}]{1995PASP..107..803U}
{Urry} C.~M.,  {Padovani} P.,  1995, \mn@doi [\pasp] {10.1086/133630}, \href
  {https://ui.adsabs.harvard.edu/abs/1995PASP..107..803U} {107, 803}

\bibitem[\protect\citeauthoryear{Viana}{Viana}{2016}]{Via16}
Viana F.~A.,  2016, Quality and reliability engineering international, 32, 1975

\bibitem[\protect\citeauthoryear{{Wallace} \& {Sarin}}{{Wallace} \&
  {Sarin}}{2025}]{WS25}
{Wallace} W.~F.,  {Sarin} N.,  2025, \mn@doi [\mnras] {10.1093/mnras/staf623},
  \href {https://ui.adsabs.harvard.edu/abs/2025MNRAS.539.3319W} {539, 3319}

\makeatother
\end{thebibliography}

\appendix

\section{Parameter posterior for TXS 0506+059 and PKS 0735+178}

We show in this appendix the parameter posterior distributions of TXS 0506+059 in Figures \ref{fig:posterior_TXS_spec} and \ref{fig:posterior_TXS} and PKS 0735+178 in Figure \ref{fig:posterior_PKS0735}, respectively.

\begin{figure*}
    \centering
    \includegraphics[width=0.95\textwidth]{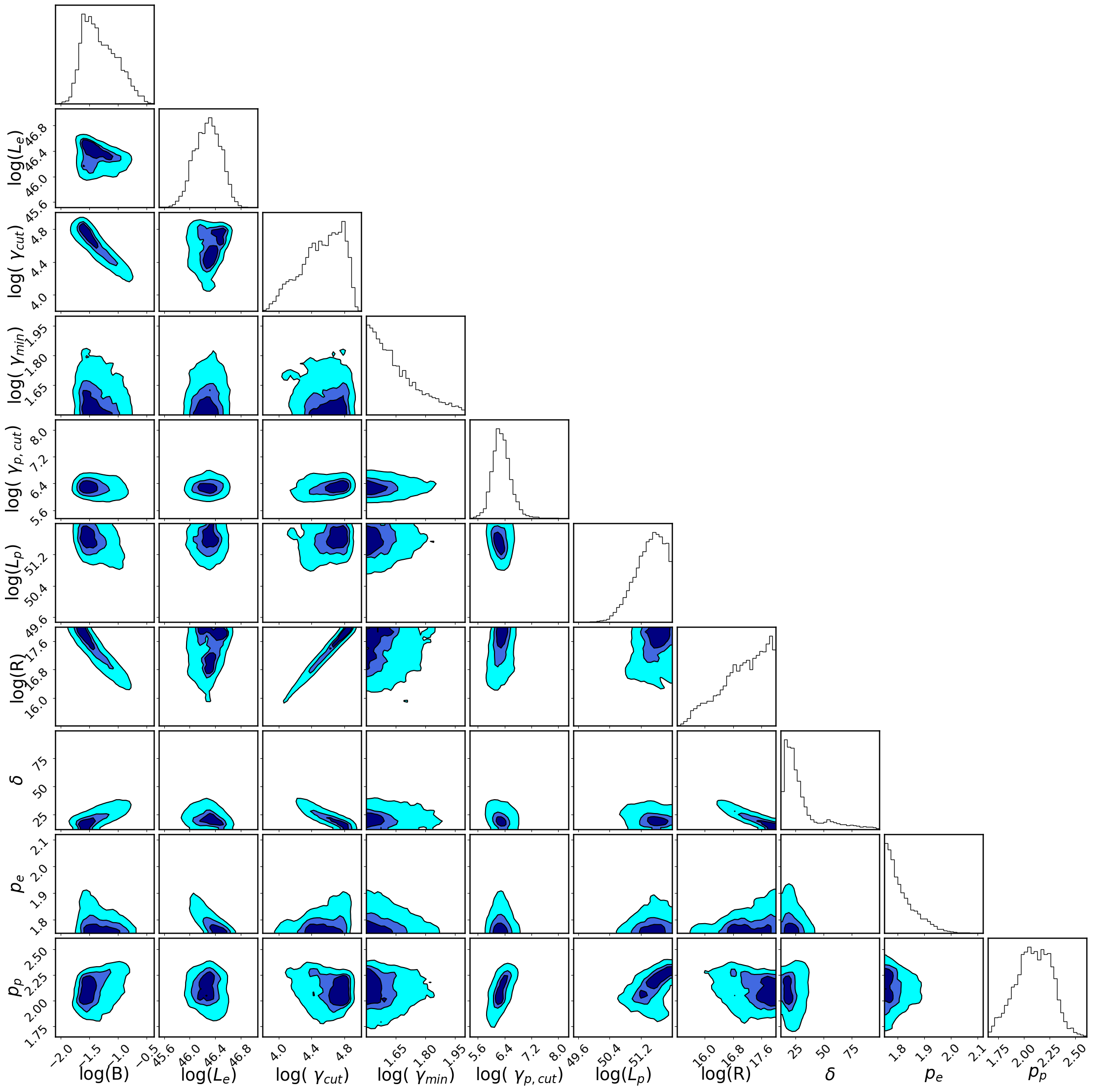}
    \caption{Posterior distributions of model parameters for TXS 0506+059 during the observation of the IceCube-170922A event, assuming an $E^{-2}_{\nu}$ neutrino spectrum with a flux of $\sim10^{-12}\:{\rm erg\:cm^{-2}\:s^{-1}}$ between energies 183 TeV and 4.3 PeV.}
    \label{fig:posterior_TXS_spec}
\end{figure*}

\begin{figure*}
    \centering
    \includegraphics[width=0.95\textwidth]{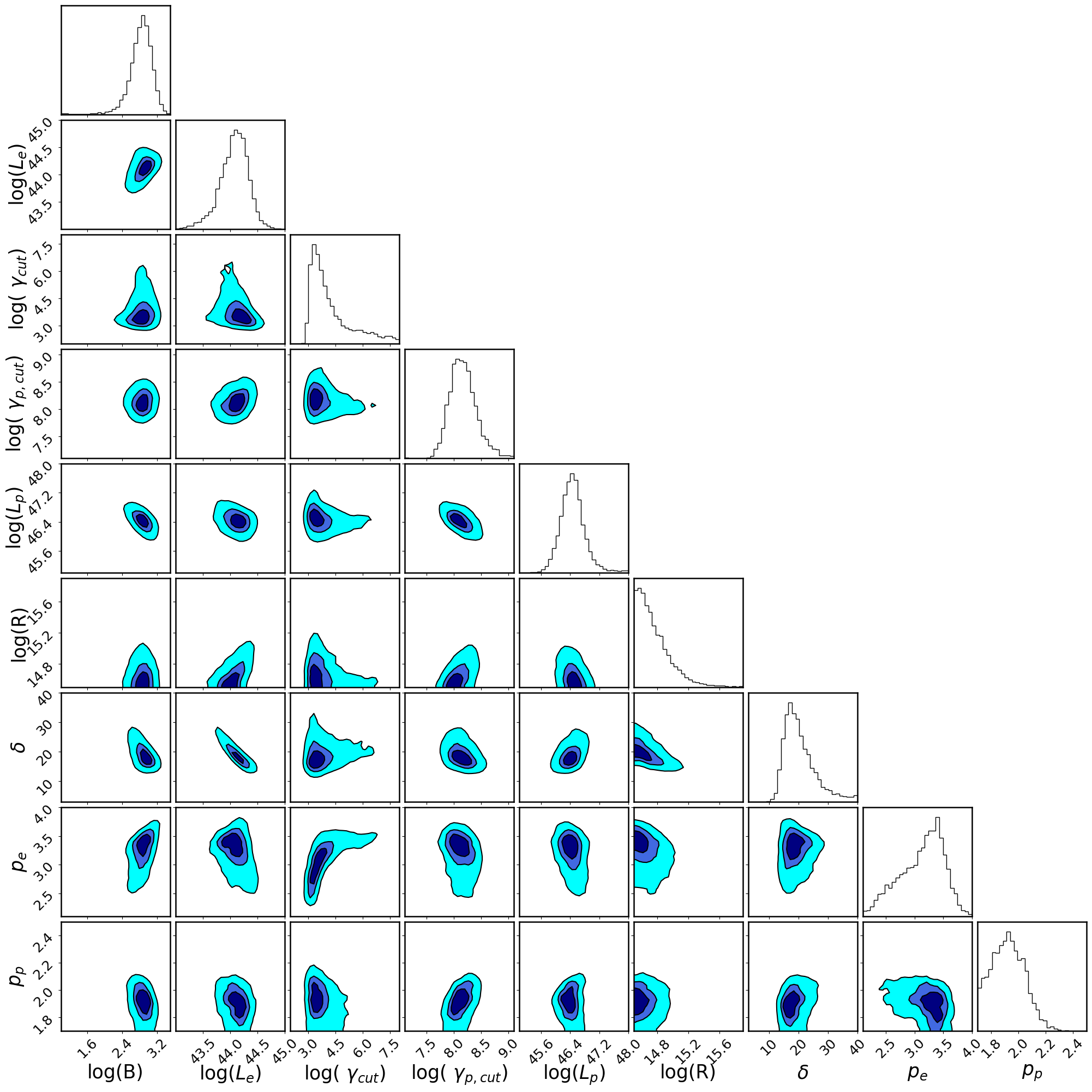}
    \caption{Same as in Figure \ref{fig:posterior_TXS_spec}, but using a Poisson likelihood approach and assuming the detection by IceCube of a neutrino event in a one-year long exposure period. In addition, $\gamma_{\rm e,min} = 100$ was assumed.}
    \label{fig:posterior_TXS}
\end{figure*}

\begin{figure*}
    \centering
    \includegraphics[width=0.95\textwidth]{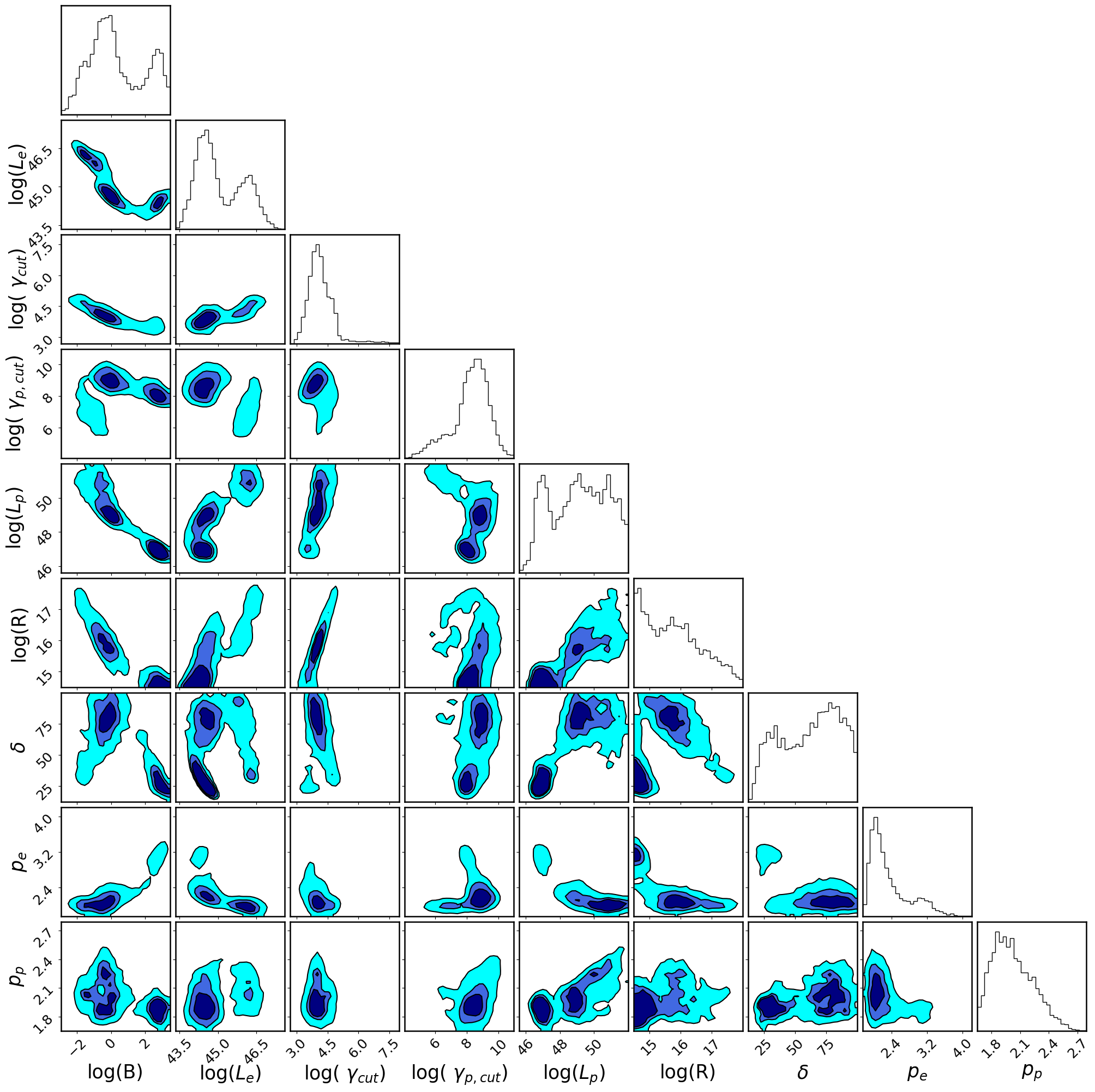}
    \caption{Posterior distributions of model parameters for PKS 0735+178 during its multiwavelength flaring period
    coinciding with the observation of IceCube-211208A. The fit was performed using a Poisson likelihood for the neutrino and
    assuming the detection by IceCube of one neutrino event over a period of one year. In addition, we set $\gamma_{\rm e,min} = 100$. Clearly, the resulting distributions are bimodal with the parameters characteristics of hydrid or P-syn models, underlying that the data used in this analysis are not sufficient to differentiate between these two models.}
    \label{fig:posterior_PKS0735}
\end{figure*}


\end{document}